\documentclass[twocolumn]{aastex63}

\graphicspath{{./}{figures/}}
\usepackage{amsmath,amsfonts,amssymb}
\usepackage{graphicx}
\usepackage{xspace}
\usepackage{upgreek}
\usepackage{amsmath,amsfonts,amssymb}
\usepackage{graphicx}
\usepackage{color, colortbl} 
\usepackage{booktabs} 
\usepackage{multirow} 
\usepackage[inline]{enumitem} 

\usepackage{wasysym}
\usepackage{physics} 
\usepackage{textcomp,gensymb} 
\usepackage{savesym}
\savesymbol{tablenum}
\usepackage{siunitx}
\restoresymbol{SIX}{tablenum}

\renewcommand{\d}[2]{\dfrac{\partial #1}{\partial #2}} 
\newcommand{\D}[2]{\dfrac{d #1}{d #2}} 
\def\arcmin{\hbox{$^\prime$}\xspace} 
\def\arcsec{\hbox{$^{\prime\prime}$}\xspace} 

\begin{document}

\title{A study of 90~GHz dust emissivity on molecular cloud and filament scales}

\correspondingauthor{Ian Lowe}
\email{ianlowe@sas.upenn.edu}

\author[0000-0003-4063-2646]{Ian Lowe}
\affiliation{Department of Physics and Astronomy, University of
Pennsylvania, 209 South 33rd Street, Philadelphia, PA, 19104, USA}

\author[0000-0002-8472-836X]{Brian Mason}
\affil{NRAO—National Radio Astronomy Observatory, 520 Edgemont Road, Charlottesville, VA 22903, USA}

\author{Tanay Bhandarkar}
\affiliation{Department of Physics and Astronomy, University of
Pennsylvania, 209 South 33rd Street, Philadelphia, PA, 19104, USA}

\author{S. E. Clark}
\affil{Institute for Advanced Study, 1 Einstein Drive, Princeton, NJ 08540, USA}

\author[0000-0002-3169-9761]{Mark Devlin}
\affil{Department of Physics and Astronomy, University of
Pennsylvania, 209 South 33rd Street, Philadelphia, PA, 19104, USA}

\author[0000-0002-1940-4289]{Simon R.\ Dicker}
\affiliation{Department of Physics and Astronomy, University of
Pennsylvania, 209 South 33rd Street, Philadelphia, PA, 19104, USA}

\author{Shannon M.\ Duff}
\affil{National Institute of Standards and Technology, 325 Broadway Boulder, CO, 80305, USA}

\author[0000-0001-7594-8128]{Rachel Friesen}
\affil{Department of Astronomy \& Astrophysics, University of Toronto, 50 St. George St., Toronto, ON M5S 3H4, Canada}

\author[0000-0001-5397-6961]{Alvaro Hacar}
\affil{University of Vienna, Department of Astrophysics, Türkenschanzstrasse 17, A-1180 Vienna, Austria}

\author[0000-0001-7449-4638]{Brandon Hensley}
\affil{Princeton University, Jadwin Hall, Washington Road, Princeton, NJ 08544}

\author[0000-0003-3816-5372]{Tony Mroczkowski}
\affil{ESO—European Southern Observatory, Karl-Schwarzschild-Str. 2, D-85748 Garching b. München, Germany}

\author[0000-0002-4478-7111]{Sigurd~Naess}
\affil{Center for Computational Astrophysics, Flatiron Institute, New York, NY, USA 10010}

\author[0000-0001-5725-0359]{Charles Romero}
\affil{Green Bank Observatory, P.O. Box 2, Green Bank, WV 24944, USA}

\author{Sarah Sadavoy}
\affil{Department for Physics, Engineering Physics and Astrophysics, Queen's University, Kingston, ON, K7L 3N6, Canada}

\author{Maria Salatino}
\affil{Kavli Institute for Astroparticle Physics and Cosmology (KIPAC)
 94305 Stanford CA, US}

\author[0000-0003-0167-0981]{Craig Sarazin}
\affil{Department of Astronomy, University of Virginia, 530 McCormick Rd., Charlottesville, VA 22904, USA}

\author{John Orlowski-Scherer}
\affiliation{Department of Physics and Astronomy, University of
Pennsylvania, 209 South 33rd Street, Philadelphia, PA, 19104, USA}

\author{Alessandro Schillaci}
\affil{Department of Physics, California Institute of Technology, Pasadena, CA 91125, USA}

\author[0000-0001-6903-5074]{Jonathan Sievers}
\affil{McGill University, 3600 University Street, Montreal, QC H3A 2T8, Canada}

\author[0000-0002-5812-9232]{Thomas Stanke}
\affil{ESO—European Southern Observatory, Karl-Schwarzschild-Str. 2, D-85748 Garching b. München, Germany}

\author[0000-0003-2300-8200]{Amelia Stutz}
\affil{Departmento de Astronomía, Facultad Ciencias Físicas y Matemáticas, Universidad de Concepción, Av. Esteban Iturra s/n Barro Universitario, Casilla 160-C, Concepción, Chile}

\author[0000-0001-5112-2567]{Zhilei Xu}
\affil{MIT Kavli Institute, Massachusetts Institute of Technology, 77 Massachusetts Avenue, Cambridge, MA 02139, USA}


\begin{abstract}
    Recent observations from the MUSTANG2 instrument on the Green Bank Telescope have revealed evidence of enhanced long-wavelength emission in the dust spectral energy distribution (SED) in the Orion Molecular Cloud (OMC) 2/3 filament on 25\arcsec (0.1~pc) scales. Here we present a measurement of the SED on larger spatial scales (map size 0.5-3\degree or 3-20~pc), at somewhat lower resolution (120”, corresponding to 0.25~pc at 400~pc) using data from the Herschel satellite and Atacama Cosmology Telescope (ACT). We then extend the 120\arcsec-scale investigation to other regions covered in the Herschel Gould Belt Survey (HGBS) specifically: the dense filaments in the southerly regions of Orion A; Orion B; and Serpens-S. Our dataset in aggregate covers approximately 10 deg$^2$, with continuum photometry spanning from 160um to 3mm. These OMC 2/3 data display excess emission at 3mm, though less (10.9\% excess) than what is seen at higher resolution. Strikingly, we find that the enhancement is present even more strongly in the other filaments we targeted, with an average excess of 42.4\% and 30/46 slices showing an inconsistency with the modified blackbody to at least 4$\sigma$. Applying this analysis to the other targeted regions, we lay the groundwork for future high-resolution analyses. Additionally, we also consider a two-component dust model motivated by Planck results and an amorphous grain dust model. While both of these have been proposed to explain deviations in emission from a generic modified blackbody (MBB), we find that they have significant drawbacks, requiring many spectral points or lacking experimental data coverage. 
\end{abstract}
\section{Introduction}
Star forming regions within the Milky Way represent an enormous variation in scale from the size of the clouds ($\sim 10^{18}$ m) down to the individual stars formed within ($\sim 10^9$ m). Giant molecular clouds (GMC) are composed of gas and dust which collapse gravitationally to form denser regions which eventually form individual stars. This process is defined by the hierarchical nature of the structure formation as the clouds form filamentary structures \citep{Hennebelle_2012} during their collapse into dense cores. These filaments, especially the smallest and densest ones (proposed characteristic size $\sim$0.1 pc, \citet{Arzoumanian_2019}), are a particularly active area of investigation \citep{Peretto_2013, Andre_2014}. These studies include cataloguing of individual stars in the process of forming \citep{Fiorellino_2020,Konyves_2020,Polychroni_2013,Stutz_2013}, mapping the magnetic field strength and orientation within the clouds \citep{Chen_2019,Fissel_2019}, and the measuring of the thermal SEDs \citep{Schnee_2014,Sadavoy_2016,Mason_2020}. 

We focus on the submillimeter and millimeter wave emission from the clouds and internal filamentary structure. This regime is dominated by thermal emission from dust grains which are generally assumed to emit as a modified blackbody (MBB), given by
\begin{equation}
    I_{\nu, Dust} \propto \nu^{3+\beta}\frac{1}{\exp(h\nu/kT_d)-1}
    \label{eq:dust_I}
\end{equation}
where $T_d$ is the temperature of the dust grain determined by the local radiation environment, and $\beta$ is the spectral index of the dust, set by the physical properties of the grains such as their composition and size. These MBB models of dust emission agree well with observations at the submillimeter and millimeter wavelengths, with dust temperatures varying across the cloud and the spectral index typically taking on a value of $1.5 < \beta < 2.5$ \citep{Schnee_2009, Sadavoy_2013,planck_int_xlviii}. 

In addition to a generic modified blackbody spectrum, there have been recent studies of more specific models which treat the emission more in-depth than a single power-law spectrum. Models from \citet{Draine_1999} and \citet{2013_draine+hensley} take into consideration the ferromagnetic resonance properties of iron-containing grains which produced large magnetic dipole cross-sections at frequencies of 50-100~GHz. Another such model is the two-component dust model of \citet{Meisner2015} which was designed to describe the emission structure of the galactic cirrus and comprises a cold, shallow $\beta$($\sim1.6$) component in combination with a hot, sharp $\beta$($\sim2.7$) component to describe the diffuse emission. A final model which seeks to describe the behavior of the dust through a description of the particles is the amorphous grain model of \citep{Paradis_2011}. This spectrum relies on model of the emissivity of the dielectric grains over the frequency band and at different temperatures.

In the north of Orion A, we find the high line-mass Integral Shaped
Filament (ISF) out of which the Orion Nebula Cluster is forming
(e.g. \citet{Stutz_2015,Stutz_2016}). The northern portion of the ISF is
commonly denoted OMC2/3, and was the subject of study at high resolution (\citet{Schnee_2014}, hereafter S14, \citet{Sadavoy_2016}, hereafter S16). Due to its high line mass and proximity, the region is bright and
observationally accessible. Meanwhile, its status as a high mass star
and cluster forming filament, with the accompanying advanced
evolutionary stage and elevated UV radiation field, motivate the
extension of studies of the thermal dust emission to other potentially
less disturbed filaments.

The emission by dust in OMC 2/3 has a rich history of investigation. S14 took the initial measurements from the MUSTANG experiment (3.3~mm) and combined them with observations from the MAMBO (1.2~mm) instrument on the IRAM (Institut de Radioastronomie Millimetrique) 30~m telescope and NH$_3$-derived temperature maps to model the spectral index $\beta$ of the thermal dust emission. This analysis showed high emission in the 3.3~mm data and produced surprisingly shallow SEDs with $\beta \sim 0.9$, which was attributed to the potential growth of mm-sized grains in the regions of interest. S16 expanded on this analysis, including 160-500~$\mu$m data from the PACS (Photodetector Array Camera $\&$ Spectrometer) and SPIRE (Spectral and Photometric Imaging Receiver) instruments as well as 2~mm data from the GISMO (Goddard-Iram Superconducting 2-Millimeter Observer) instrument to further constrain the dust SED. These more constrained SEDs displayed a $\beta \sim 1.7-1.8$, in line with expectations (S16). The shallower $\beta$ of S14 is then attributed to enhanced emission at 3.3~mm (90~GHz).

The most recent development comes from \citep{Mason_2020}, hereafter M20, which investigated the enhanced emission found in S16. M20 includes data from the commissioning of the MUSTANG2 (Multiplexed Squid-TES Array at Ninety Gigahertz 2) instrument as well as the GBT (Green Bank Telescope) Ka band receiver at 31 GHz ($\sim$1 cm) to further extend the low frequency arm of the SED. Additionally, this analysis includes SCUBA-2 (Submillimetre Common-User Bolometer Array 2) observations at 450 and 850~$\mu m$ which serve to further constrain the low-frequency tail of the SED. Results from M20 indicate that the enhanced emission seen at 3.3~mm is confirmed by the 1~cm data point and that it is inconsistent with both anomalous microwave emission (AME) and spinning dust emission. Such flattening of the SED has also been predicted theoretically from amorphous dust models \citep{Meny_2007,Paradis_2011,Coupeaud_2011,nashimoto2020}, but these predictions have not yet been confirmed observationally.

Understanding the emission properties of astrophysical dust and its spectrum is important for much of astrophysics and cosmology. Dust emission is often used as a mass tracer and estimator \citep{eales_2012,Groves_2015,Paradis_2019} and a systematic disagreement with models would significantly affect dust mass estimates. In addition, polarized emission is used to trace the plane-of-the-sky projected magnetic field and, under specific assumptions, estimate the field strengths \citep{Heitsch_2001,Fissel_2019,crutcher_magnetic}. Pertaining to cosmology, the cosmic microwave background (CMB) is measured through sight-lines that are contaminated by the polarized emission of the dusty interstellar medium of the Milky Way. These foregrounds are modelled \citep{Vansyngel_2018,Hensley_2018} and removed to get at the underlying CMB and its polarization spectrum. These important uses of polarized dust emission and the models upon which they are based must, however, also describe the unpolarized emission, in which we have begun to see a break in the spectral properties between 150 and 90~GHz. To this end, we have expanded the studies of S14, S16, and M20 of OMC 2/3 to additional clouds within the Herschel Gould Belt Survey (HGBS) to garner an understanding of whether this break in the spectral properties is widespread in nearby molecular clouds. In this paper we use the methods of M20 extended to larger angular scales and other regions using data from the Atacama Cosmology Telescope (ACT), Herschel PACS, and Herschel SPIRE.

The structure of the paper is as follows. In Section \ref{sec:data} we review the data used in each analysis mode (25\arcsec and 120\arcsec), discuss the regions we have targeted in this work, and lay out the map production pipeline. Section \ref{sec:analysis} details the method of SED extraction and the models used to fit the data. In Section \ref{sec:valid} we compare the results of the pipeline at 25\arcsec to the previous analysis as well as discuss potential sources of contamination within the data. We discuss the results of this study and the potential for follow-up in Section \ref{sec:results}. Finally, we conclude in Section \ref{sec:concs}. In this paper, we refer to the difference between a model-predicted brightness as enhanced emission or the elevation of a data point and all error bars represent a 1-$\sigma$ uncertainty (68$\%$ confidence) on the associated quantity.
\section{Datasets}
\label{sec:data}

The data that were used for this study varied between the high- and low-resolution analyses. In the initial high-resolution verification and calibration of the pipeline, we included data from the PACS, SPIRE, SCUBA-2, MAMBO (Max-Planck Millimeter Bolometer Array), GISMO, and MUSTANG2 instruments to cover a spectral range from 160$\mu$m to 3mm at a resolution of 25\arcsec matching the 350$\mu$m SPIRE array. The study was extended to low-resolution with the wealth of survey data provided by the ACT instrument. For the low-resolution analysis we retain the Herschel survey data on the high frequency end and combine it with the ACT full-sky data on the low-frequency side. The enormous footprint of the ACT map affords coverage of almost any HGBS region we aim to target. Within the HGBS we targeted a handful of star forming regions, particularly chosen to span a range in density and star formation activity level (isolated {\it vs} clustered) while retaining a molecular hydrogen column density in excess of 1e22 cm$^{-2}$ \citep{Polychroni_2013,Konyves_2020,Fiorellino_2020}. For these low-resolution analyses, we match the ACT 90~GHz instrument which has a beam size of 120\arcsec and cover the same spectral range of 160$\mu$m to 3~mm.

\subsection{Targeted high-resolution photometry}
\subsubsection{MUSTANG2}
At 90~GHz, our high-resolution data comes from the MUSTANG2 instrument on the Robert C. Byrd Green Bank Telescope (GBT). MUSTANG2 features 215 horn-coupled transition edge sensor (TES) bolometers, a bandpass of 75-105~GHz, a field of view of 4.25\arcmin, and a beam size of 10\arcsec giving excellent resolution and mapping power at 90~GHz. For more information about the MUSTANG2 instrument, see \citet{dicker_2014} and \citet{stanchfield_2016}.

In addition to producing the initial OMC 2/3 90~GHz map, this instrument provides an excellent option for high-resolution followup mapping of interesting regions seen in the larger survey analysis. As the excess signal is seen in the low-frequency (90~GHz and longer $\lambda$) tail, the availability of a high-resolution and on-sky resource for mapping these regions is important to the understanding of this behavior.
\subsubsection{Legacy Datasets}
Our data in the 160, 250, and 350 $\mu$m bands come from the HGBS observations with the PACS and SPIRE instruments. These bolometer cameras operated on the Herschel satellite \citep{Pilbratt_2010}. The 350$\mu m$ band sets the limiting resolution of 25\arcsec to which we match all of the maps. At 450 and 850 $\mu m$ our data comes from the SCUBA-2 instrument \citep{Holland_2013} which is a dual-band bolometer receiver on the James Clerk Maxwell Telescope. The remaining data at 1.2mm and 2mm come from the GISMO \citep{staguhn_2006} and MAMBO \citep{Bertoldi_2003} instruments which operated on the IRAM 30m telescope.

\subsection{Wide Area Survey Data}
\subsubsection{ACT}
The Advanced ACTPol instrument is a cosmic microwave background (CMB) camera on the Atacama Cosmology Telescope that observes roughly half of the sky, mapping the polarization spectrum of the CMB and measuring the Sunyaev-Zel'dovich Effect (SZE) of galaxy clusters. The experiment features more than 5500 polarized transition edge sensors (TES) in five bands, centered at nominal frequencies of 28, 41, 90, 150, and 230~GHz. Further information about the AdvACT experiment can be found in \citet{Henderson_2016}.
We use the 90-230~GHz data as the low-frequency data has not yet been analyzed, which sets a resolution limit of 120\arcsec due to the beam size of the 90~GHz band. For this analysis, we used the data release which included observations through the 2017--2018 season \citep{Naess_2020} which has been combined with the Planck maps \citep{Planck2013_1} to recover large scale flux around bright sources such as OMC 1. The product used here is cutouts of the HGBS regions from the full ACT map. Notably, this 3~mm data product is calibrated using the CMB, rather than the planets used for MUSTANG2 observations, which provides an additional layer of insulation from systematics that might contaminate our conclusions.
\subsubsection{SPIRE}
Extending the scope of this analysis to 120\arcsec we give up access to the 450 and 850~$\mu$m spectral coverage provided by the SCUBA-2 instrument due to severe filtering effects. We recover this region of the spectrum through the inclusion of the SPIRE 500~$\mu$m HGBS maps, which, with a native resolution of 36.7\arcsec, no longer exceed the resolution limit of 25\arcsec set by the 350~$\mu$m maps in the high-resolution analysis. This coverage at 500~$\mu$m provides an important data point in the center of the SEDs between the 350~$\mu$m SPIRE and 1.4~mm (220~GHz) ACT data in the transition into the low frequency tail.

\begin{table}
    \centering
    \begin{tabular}{|c|c|}
    \hline 
       Instrument (Band)  &  $\nu_0$ (GHz) \\
       \hline
       \textit{PACS} (160~$\mu$m)  & 1910.7 \\
       \hline
       \textit{SPIRE} (250~$\mu$m)  & 1221.8 \\
       \hline
       \textit{SPIRE} (350~$\mu$m)  & 871.9 \\
       \hline
       \textit{SPIRE} (500~$\mu$m)  & 612.28 \\
       \hline
       \textit{ACT} (1360~$\mu$m)  & 224.75 \\
       \hline
       \textit{ACT} (2000~$\mu$m)  & 148.33 \\
       \hline
       \textit{ACT} (3300~$\mu$m)  & 96.87 \\
       \hline
    \end{tabular}
    \caption{Response-weighted average frequencies of each of the instruments used in this analysis.  In most cases, this amounts to a roughly $1\%$ change in the nominal band-center, with the ACT 90GHz band seeing the largest shift due to a particularly high-frequency weighted filter.}
    \label{tab:bandpass}
\end{table}

\subsection{Instrumental Bandpasses}
\label{subsec:bandpasses}
To ensure our data on the spectral emission curves are properly located in frequency space, we calculate the response-weighted average frequency of each instrument. For the Herschel instruments, we use the frequencies calculated in M20, and, for the ACT instrument, we obtained the spectral response curves $R(\nu)$ and calculated the band-center as
\begin{equation}
    \nu_0 = \frac{\int R(\nu)\nu d\nu}{\int R(\nu)d\nu}.
\end{equation}
The response-weighted band centers can be found in Table \ref{tab:bandpass}. These bandpass-corrected frequencies are used in the analyses performed in Section \ref{subsec:SED} and amount to small shifts (typically $\sim 1\%$) in the spectral location of the data points.

\section{Data Reduction}
\label{sec:analysis}

In this section we detail the methods and tools used extract the SEDs from the original maps. We present the pipeline that we have developed to bring these disparate data products into a usable and common format. The discussion of the pipeline includes also the testing and verification steps taken to ensure that the data are treated properly. Throughout this analysis, all flux data are associated with the errors found in \ref{tab:errors}, which includes calibration error (Herschel - \citet{Bendo_2013}, ACT - \citet{Naess_2020}), typical noise levels in the maps, and the noise introduced through the map processing steps.

\begin{table}
    \centering
    \begin{tabular}{|c|c|c|c|}
    \hline
       Instrument & Pipeline & Map Noise & Calibration \\ \hline
       ACT 90 & $<$1\%  & 4.3\% &  4\% \\ \hline
       ACT 150 & $<$1\%  & 2.8\% &  4\% \\ \hline
       ACT 220 & $<$1\%  & 2.2\% &  4\% \\ \hline
       SPIRE 500 & $<$1\%  & 1.8\% &  5.5\% \\ \hline
       SPIRE 350 & $<$1\%  & 2.0\% &  5.5\% \\ \hline
       SPIRE 250 & $<$1\%  & 2.4\% &  5.5\% \\ \hline
       PACS 160 & $<$1\%  & 2.2\% &  5\% \\ \hline
    \end{tabular}
    \caption{Associated errors for each instrument and, in the case of the specific map noise, each specific band. The map noise levels are the rms noise of the signal levels seen in a roughly 82 kilopixel region in each map multiple beam distances from the filamentary structures as compared to the typical signals at each peak.}
    \label{tab:errors}
\end{table}




\begin{figure*}[htp]
    \centering
    \includegraphics[scale=0.55]{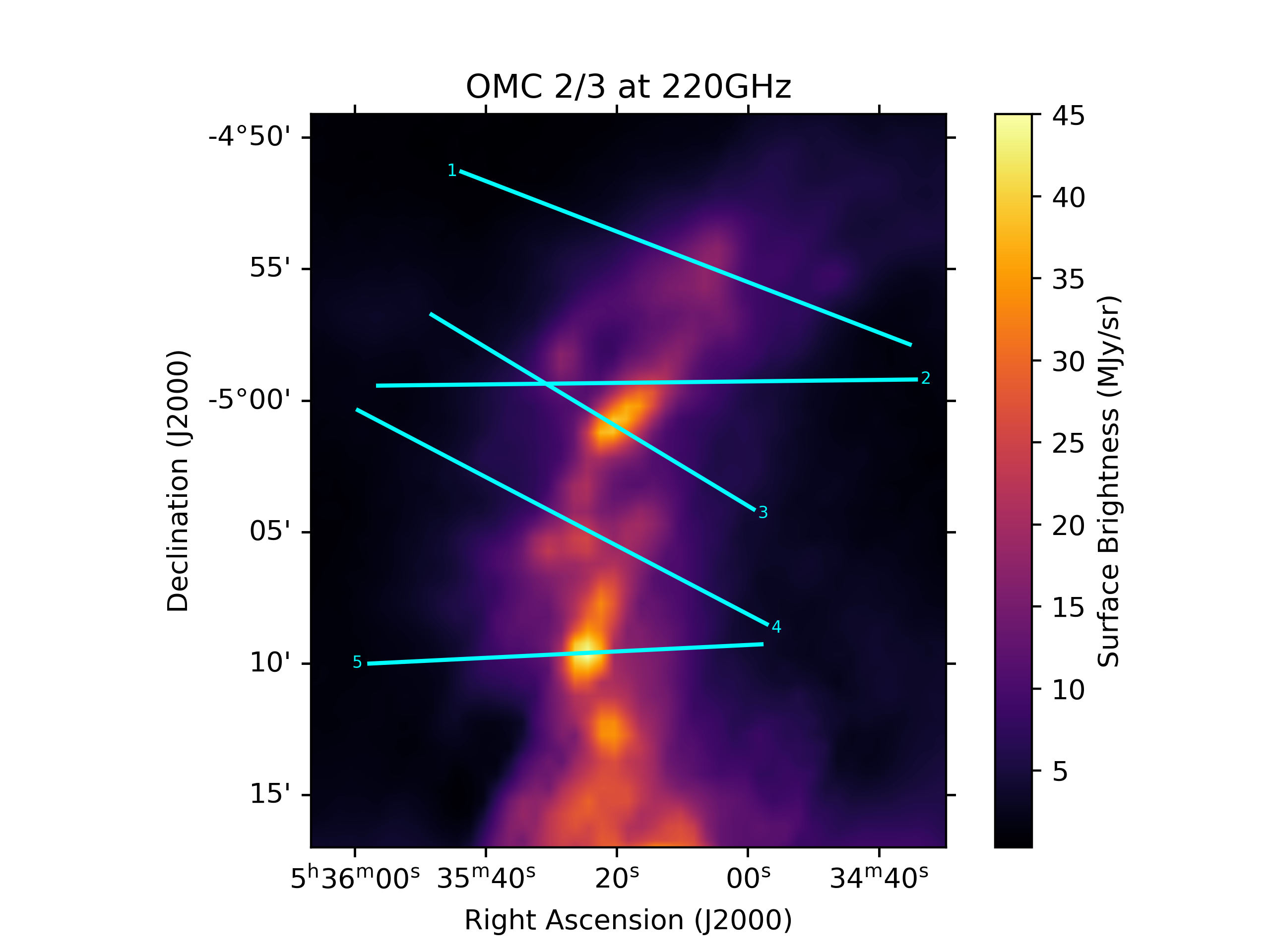}
    \includegraphics[scale=0.55]{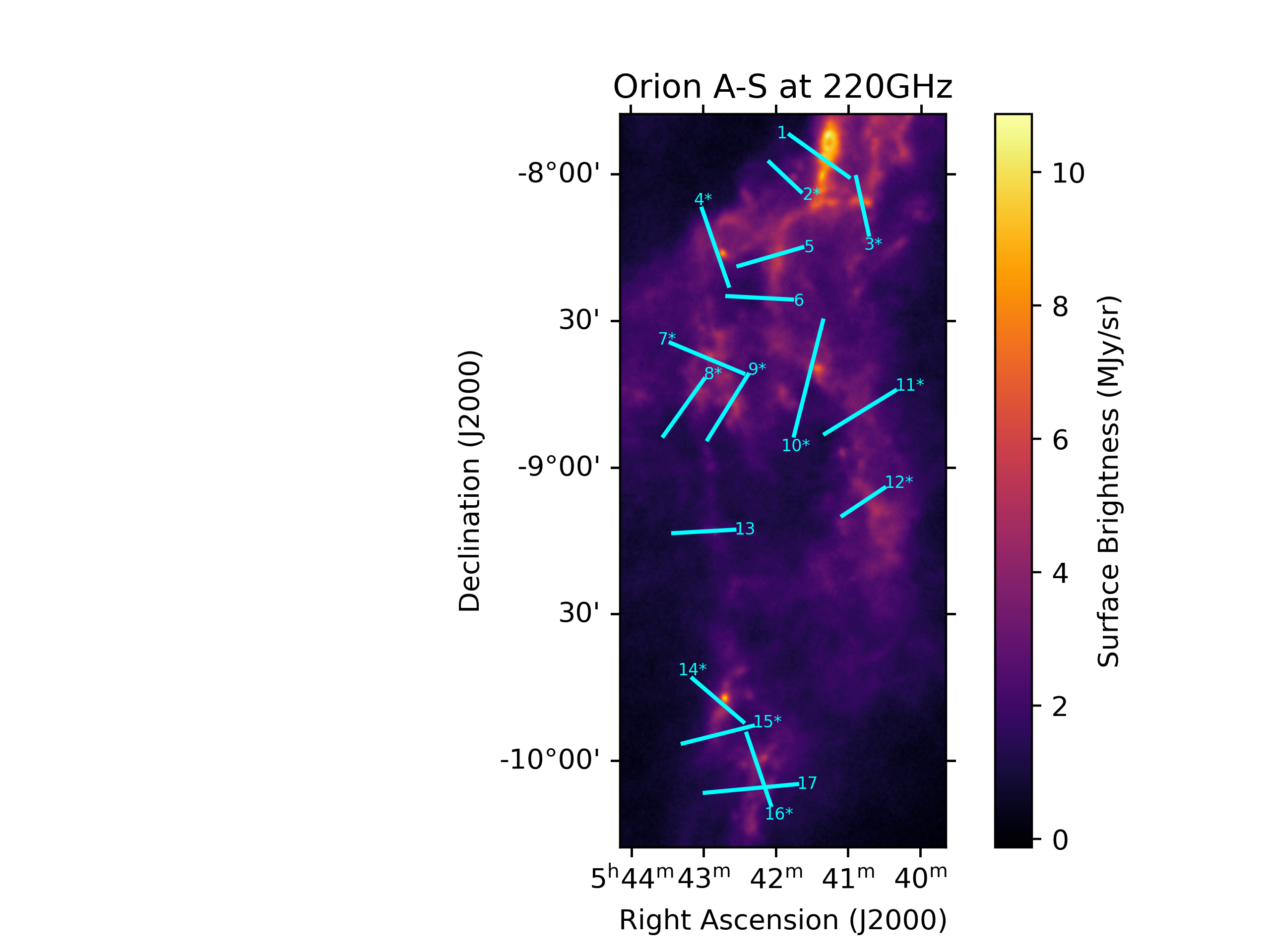}
    \includegraphics[scale=0.55]{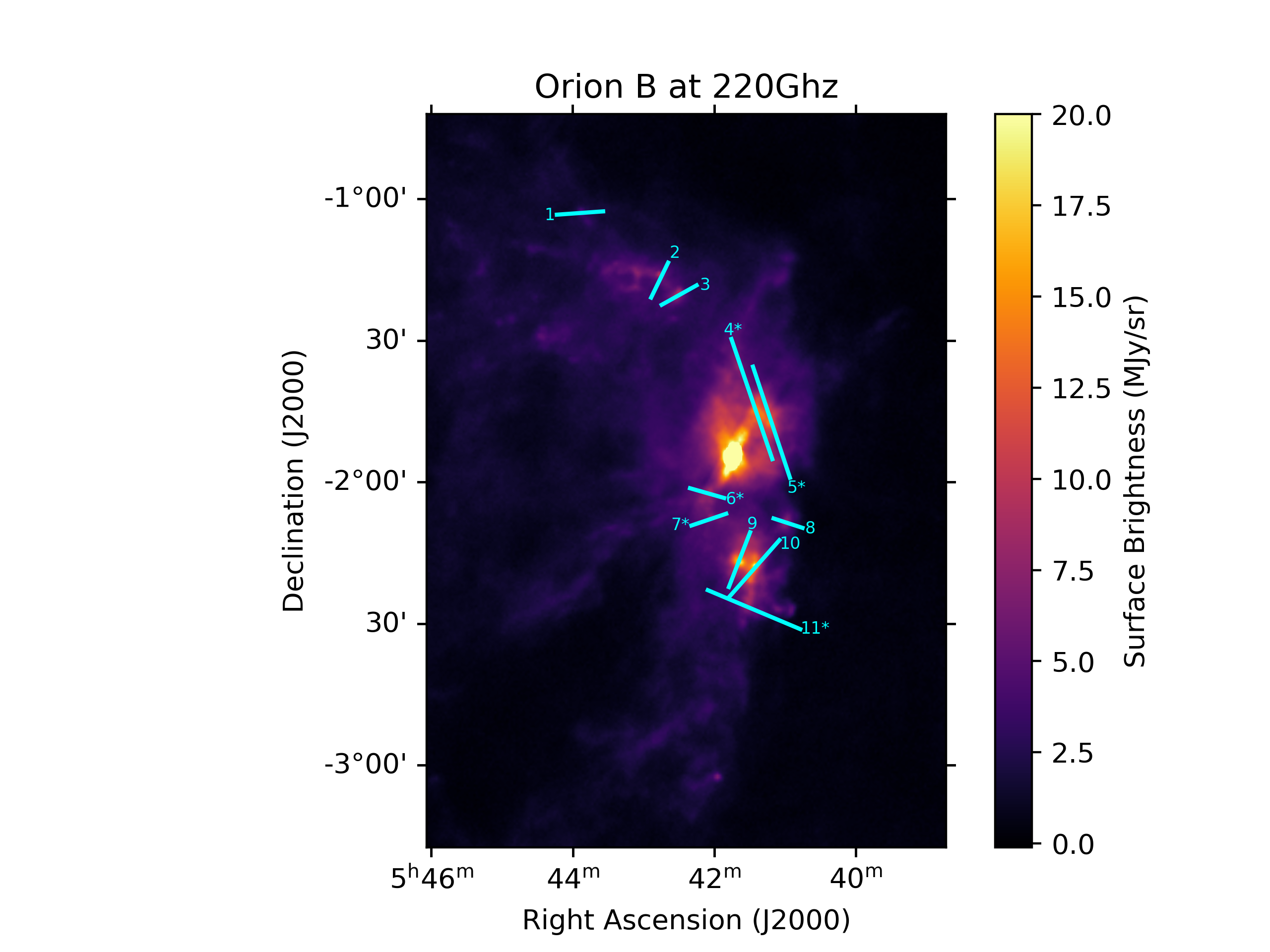}
    \includegraphics[scale=0.55]{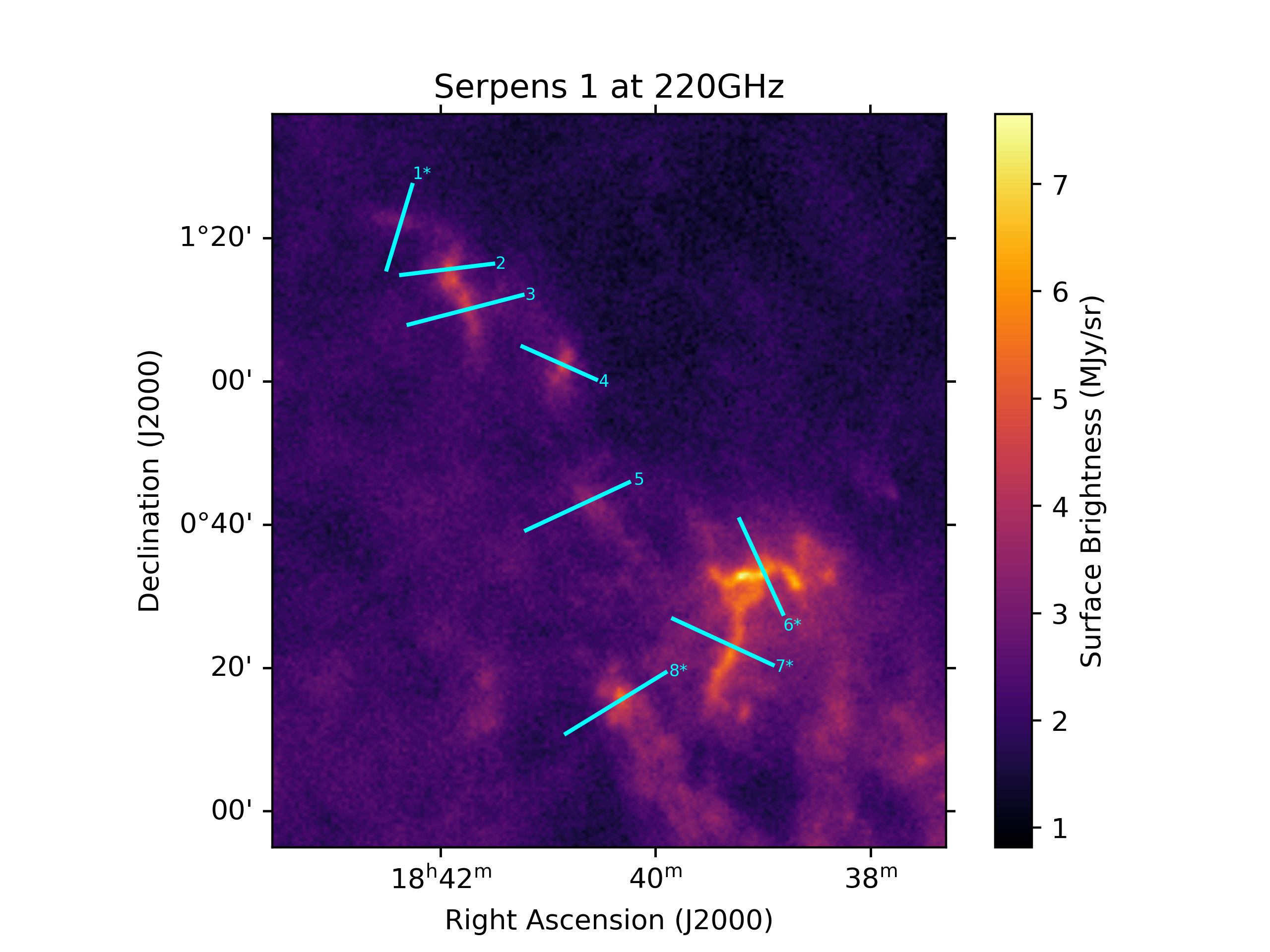}
    \includegraphics[scale=0.55]{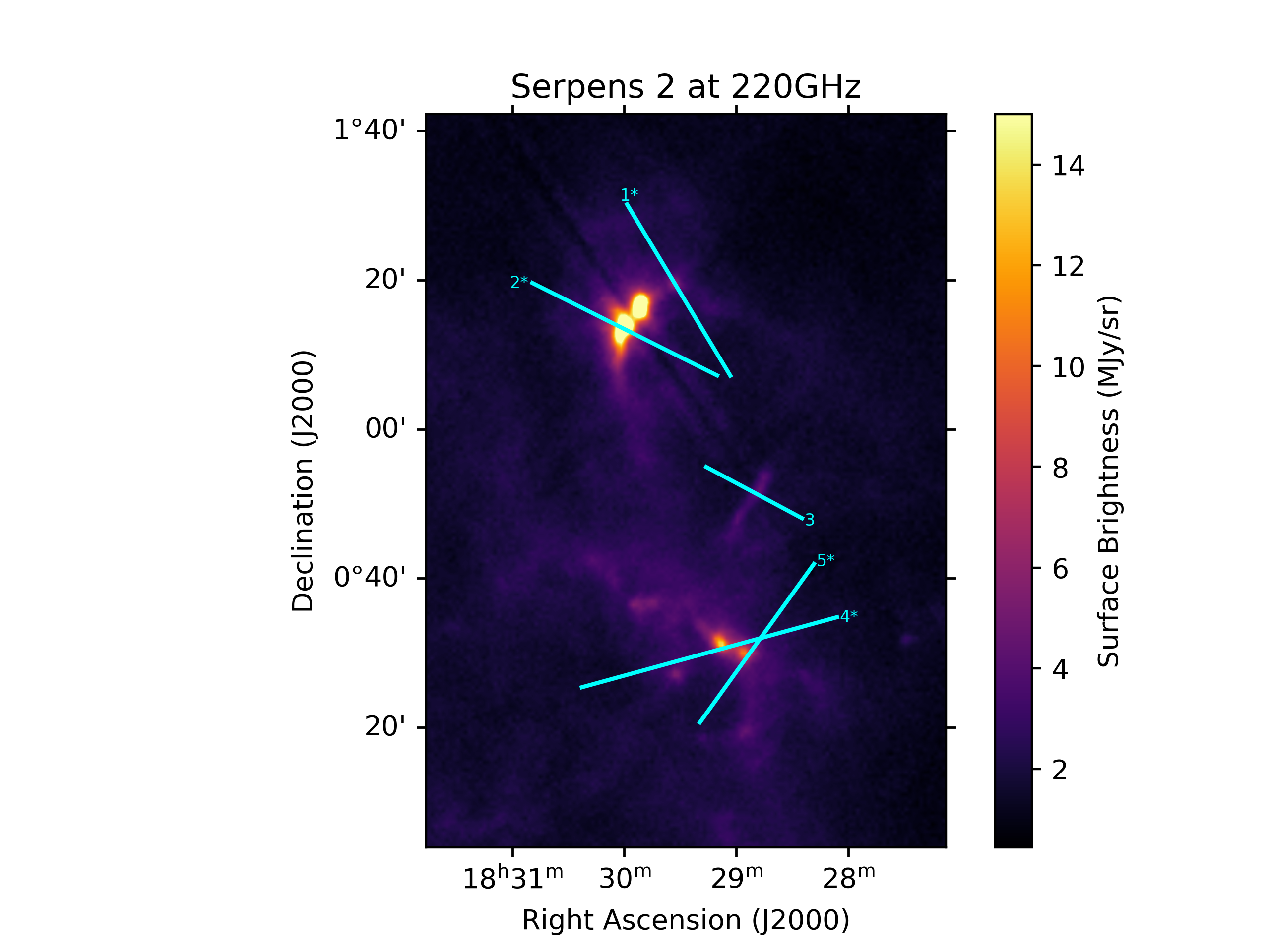}
    \includegraphics[scale=0.55]{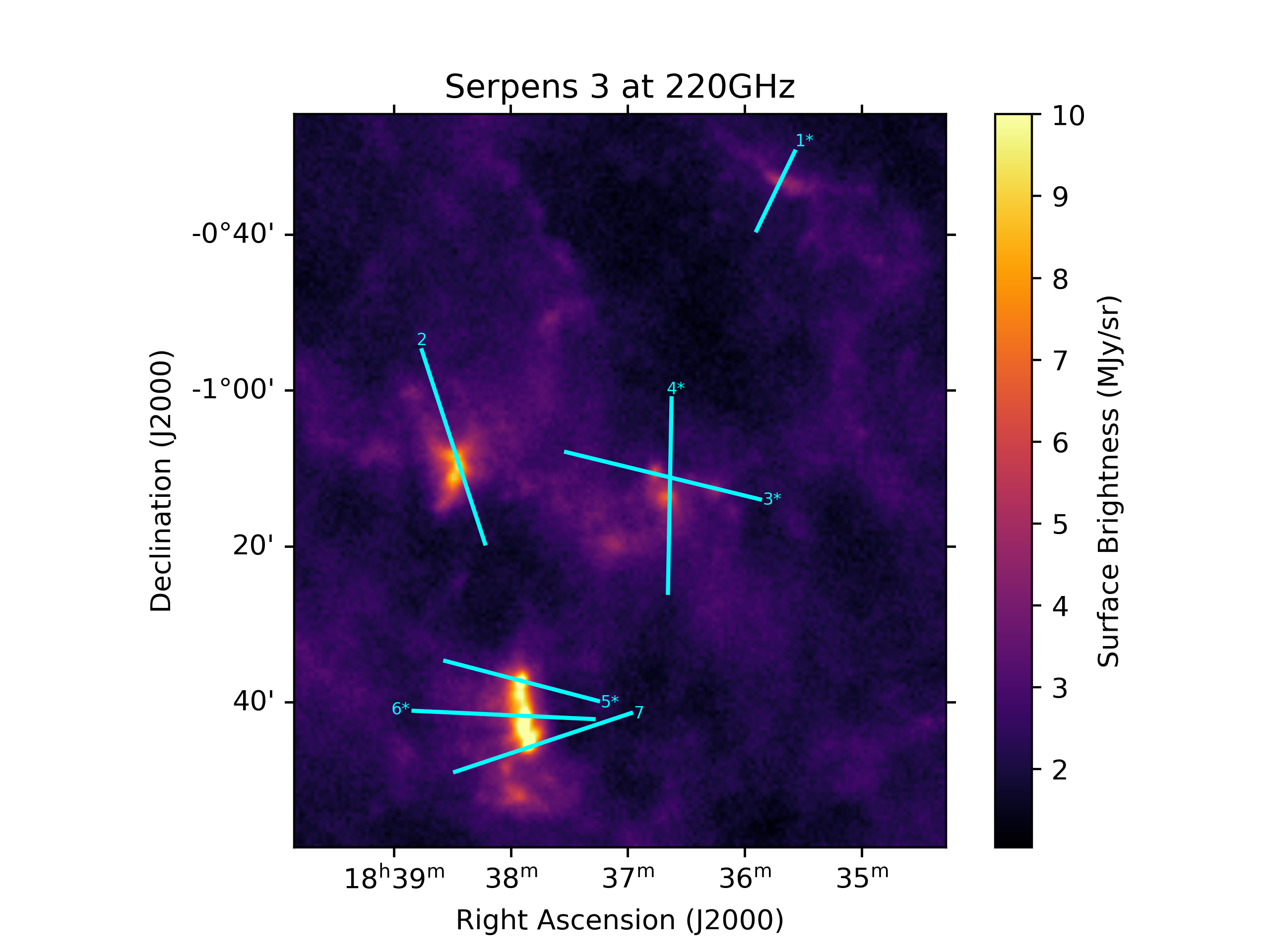}
    \caption{The six regions examined in this analysis with the selected slices overlaid on the ACT 220~GHz maps. Slices marked with stars showed excess 90~GHz emission at at least a 4-$\sigma$ significance.}
    \label{fig:slices}
\end{figure*}

\subsection{Pipeline}
A goal of this analysis was to recreate and verify the data reduction pipeline in python based on the \texttt{IDL}-produced results from M20. As such, we started with a goal of accurately reproducing the results seen in the 25\arcsec analysis of the OMC 2/3 region, while creating a scalable framework that could be expanded to include other data, such as the 120\arcsec data presented in this paper, with minimal modification. We break the process down into multiple steps, each of which we discuss below.
\subsubsection{Initial maps}
The first step was to extract sub-maps around the regions of interest from much larger maps, such as the full-sky ACT maps or large Herschel regions. The rough postage-stamping is achieved through a script which uses the functionality of the \texttt{astropy.io} library to load the ACT maps and another map as a set of reference coordinates, which are then used to create a preliminary, large rectangular cutout of the region. 


Once the maps have been postage stamped, the \texttt{fits} files are then scanned using a wrapper function which parses all maps' \texttt{bunit} keywords, compares them to a dictionary of known units, and converts them to the desired output unit (MJy/sr in this analysis). 
Once all maps have been properly resized and the values correctly mapped in the common and desired units, they are saved in preparation for the next step.

\subsubsection{Reprojected maps}
The next stage in the pipeline is to convert all maps to the same world coordinate system (WCS), pixelization, and desired region. Our output WCS is defined by first selecting the targeted region in a map in DS9 and determining the size and central pixel coordinates from the initial maps. The original map coordinates are then scaled to the new pixelization and an output shape along with a WCS header is created for this selected region. Once these objects are created, we resample the maps to the new coordinate system using the \texttt{reproject} package which is an extension of the utilities found in the \texttt{astropy} library. Specifically, we use the \texttt{reproject$\_$interp} function, which performs a bilinear resampling of the original \texttt{fits} image and WCS onto the newly defined WCS. For these maps, we chose to use a final pixelization of 2\arcsec to match that of the original MUSTANG2 map.

We performed extensive verification of these functions in order to verify that the output data is both properly scaled and in the correct locations. The first order test mapped the original data to the same WCS, which showed that the manipulations did not shift the locations of the data or appreciably change the values of the pixels. After verifying this, we next tested the flux conservation of the \texttt{reproject} functions, we expect that
\begin{equation}
    \Sigma p_0 = (\frac{n_{p_0}}{n_{p'}}) \Sigma p'
\end{equation}
where $p_0$ is the pixel values in the original map and p$'$ is the pixel values in the new map. The right hand side is also scaled by the relative pixel sizes to conserve flux. This test was applied to both the entire map and small test regions around bright sources in different maps to ensure that there was no bias affecting larger flux values. Results of this test showed that the effect was 0.05-0.5$\%$ in magnitude and that the exact value depended on the structure and change in pixelization of the map itself, which is included our assumed 5$\%$ map error.

\subsubsection{Resolution matching}
The final stage in the map production pipeline is to match the resolution of the maps to that of the lowest resolution instrument. Consider a data-product map, which consists of a grid of sampled data points and associated values (data) and convolved with the instrument point-source response kernel ($B_0$), creating the initial map ($map_0$). To degrade the resolution of this map to that of an instrument with a worse point source response ($B_f$), a second convolution with another response kernel ($b$) must be applied such that
\begin{equation}
    map_{f} = map_0*b = data*B_0*b = data*B_{f}.
\end{equation}
For the simplest case of a Gaussian beam, which we assume here, this can be solved for the characteristic convolution function \textit{b} as a Gaussian smoothing function obeying,
\begin{equation}
    \sigma_{b}^2 = \sigma_{final}^2-\sigma_0^2
\end{equation}
Where these $\sigma$ are the standard deviations of the Gaussian smoothing functions. Once these values have been obtained, we create a set of 2D Gaussian convolution kernels and apply them to the maps using the \texttt{astropy.convolution} library and associated functions. As the standard convolution kernel application sets the values beyond the map boundary to 0 for computational purposes, the maps were significantly oversized during development in order to prevent artefacts from appearing in the slice datasets. Once these convolutions have been applied, the maps are in their final states and are fully prepared for data extraction and SED production.

We verified the flux conservation of this method across the entire map as, excluding edge effects, the sum of the pixel values in the map should be static. For the initial 25\arcsec analysis, the fractional difference in total pixel values ranged from $10^{-8}-10^{-17}$, effectively zero, with the largest difference in the most drastically changed maps and no values causing appreciable change in the maps. At 120\arcsec, the edge effects are more important due to the larger convolution kernel and the boundary conditions. Because of these effects, we see flux conservation errors of up to a $0.5\%$ level across the entire map, though the level returns to the virtually undetectable of the 25\arcsec analysis in the central, targeted region of the map where the SED data are extracted.

\begin{figure}
    \centering
    \includegraphics[scale=0.5]{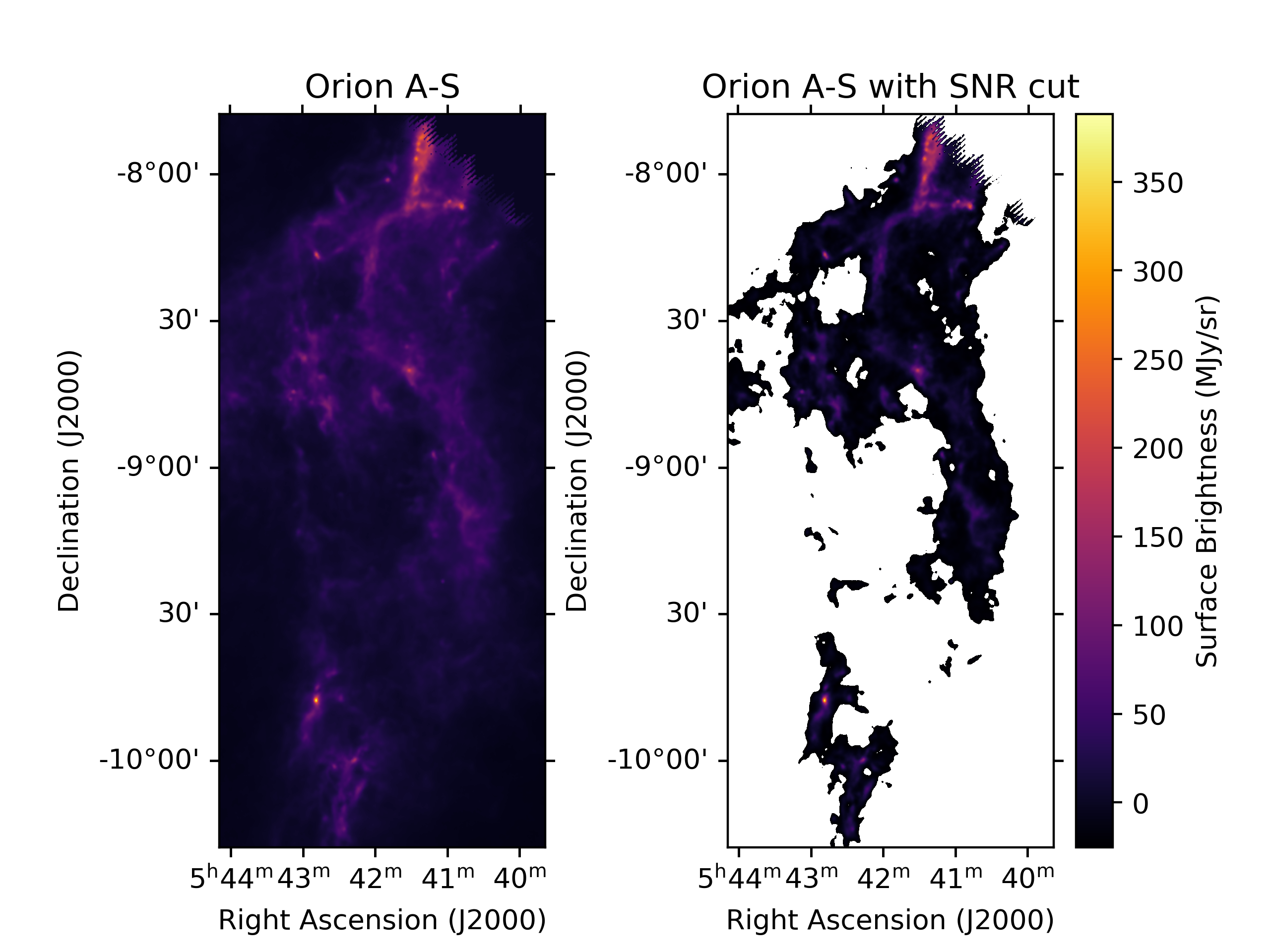}
    \caption{Example of the SNR cut performed on the OrionA-S 500~$\mu m$ map. The large white voids cover the areas below the cutoff brightness values, and slices target sections within the high signal regions.}
    \label{fig:snr_cut}
\end{figure}

\begin{figure*}[ht!]
    \centering
    \includegraphics[scale=1]{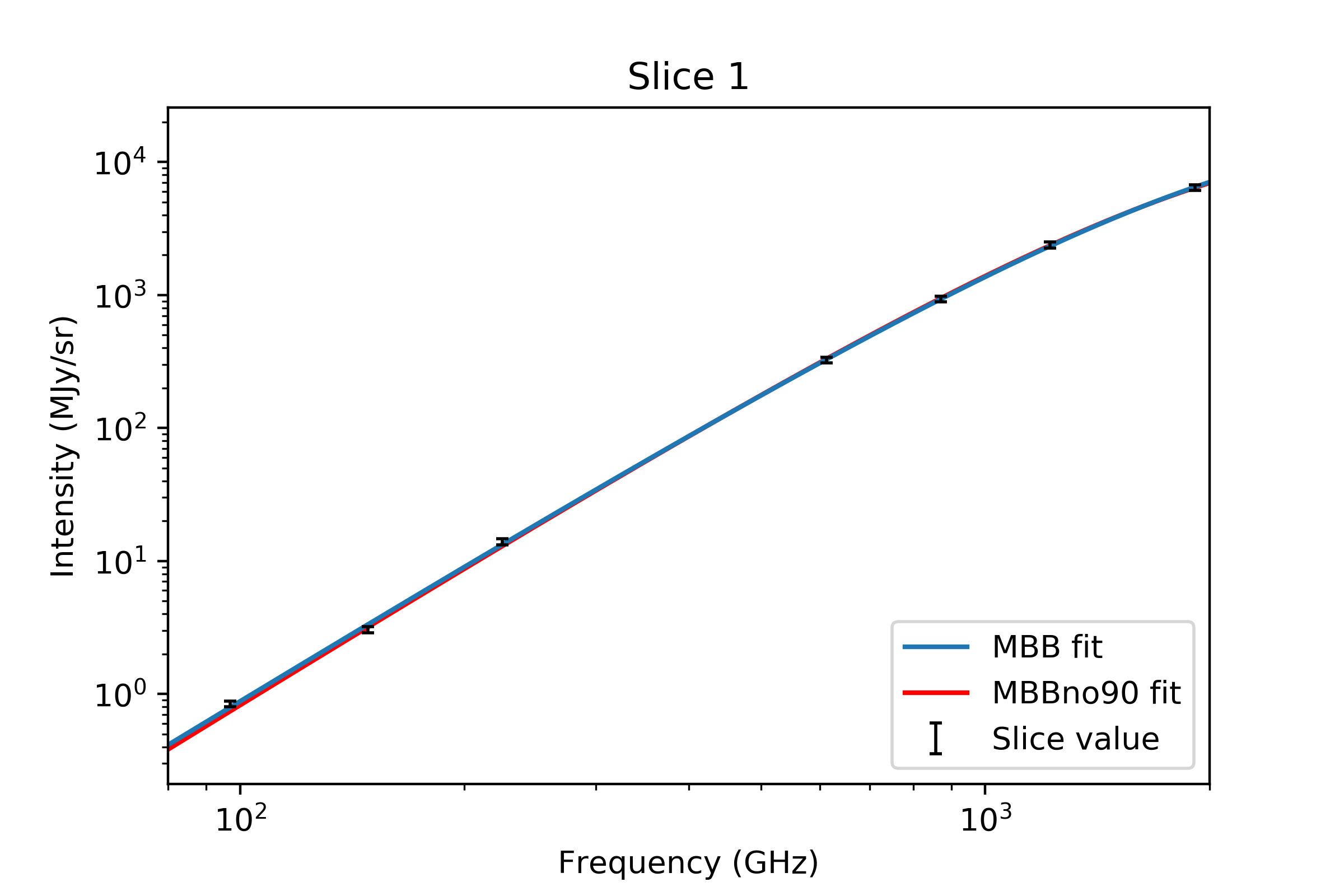}
    \includegraphics[scale=1]{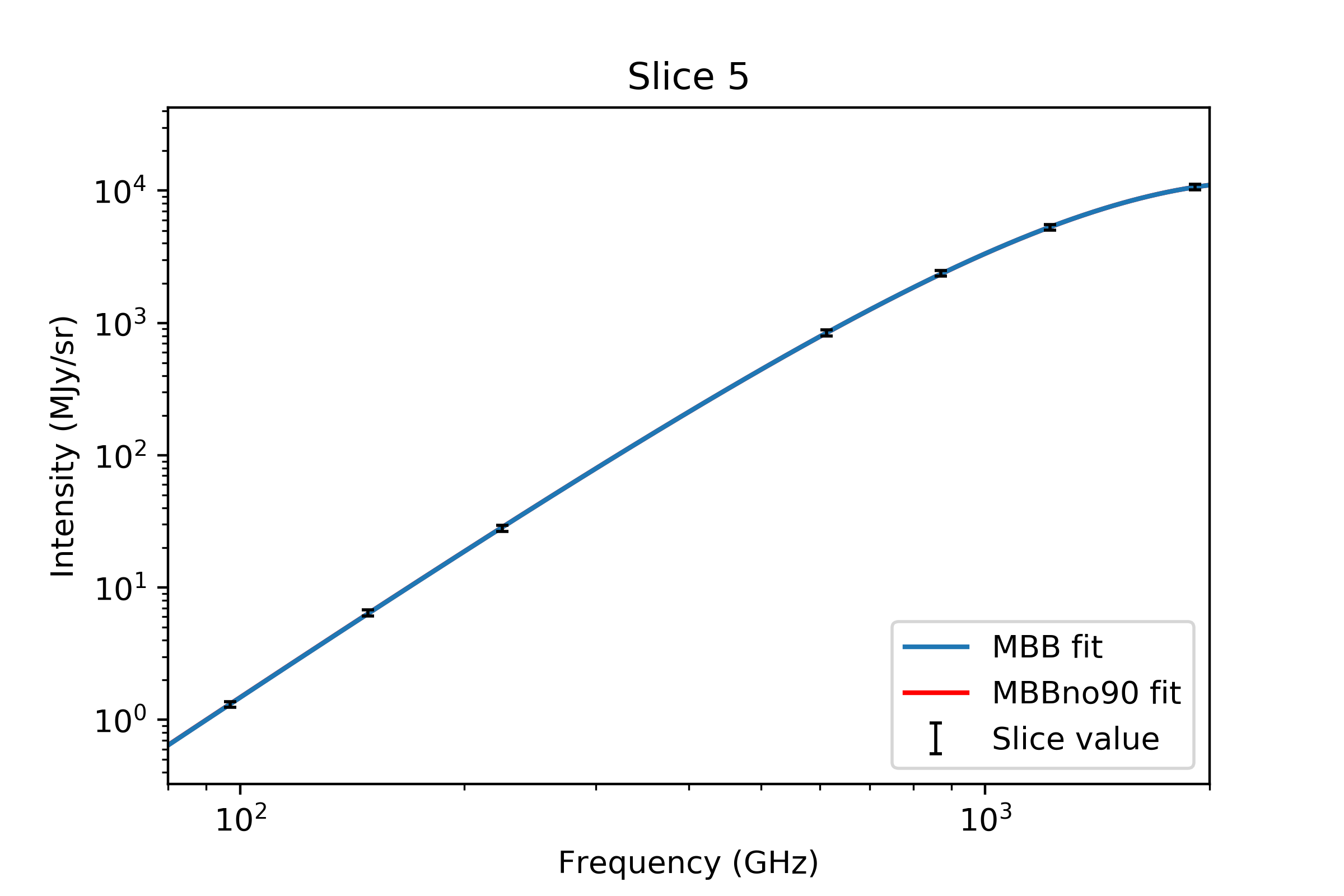}
    \caption{Spectral energy distributions for two example slices in the OMC 2/3 region at 120\arcsec resolution. The blue lines trace the \textbf{MBBall} to all data and the red the \textbf{MBBno90} fit. Additional SEDs are available upon request.}
    \label{fig:omc23-SEDs}
\end{figure*}

\subsubsection{Slices}
To extract our SED datapoints, we used a slicing and deramping method. The concept behind this method is to remove any linear drifts in the background as well as the mean levels that may enter from map making or diffuse emission not associated with the filament itself. For the 25\arcsec analysis of OMC 2/3 (verification of the pipeline), the slices were in predetermined locations from the previous analysis in M20. For all other analyses (120\arcsec investigations), the location of the slices was determined by hand and covered the high SNR filamentary structures in each map. These slices can be found for all regions in Figure \ref{fig:slices}. We perform the slicing and deramping operation as the maps are zero-median and this allows us to subtract a linear drift and remove the background mean level near the clouds. The 500~$\mu m$ maps were chosen to be the representative maps for SNR cuts as they represented both the center of the spectral band and a middle resolution. 

In order to choose slices that minimized the noise contribution of looking at low-signal areas, we first made rough signal-to-noise cuts of the regions in each 500~$\mu m$ map. To do this, we selected a low emission region far away from the filaments of approximately $10^5$ pixels in each map, calculated the mean and variance of the signal levels in these dark regions, and subtracted the mean from the map and divided by RMS noise level. With these new SNR maps, we masked all regions below a cutoff level of 5-$\sigma$ as being regions considered ``low-signal". An example of this procedure applied to the OrionA-S map can be seen in Figure \ref{fig:snr_cut}. We then imported these maps into DS9 and manually created a series of slices that extended far off of the filament into the voided areas at either end and were oriented to slice through regions of varying signal level. Once created, these slices were converted into pairs of endpoints which were fed into the data extraction algorithm detailed below.

\subsubsection{Data extraction}
\label{subsec:extraction}
The final step in the data processing pipeline is the extraction of the individual SED datapoints from each map. The data are extracted using an algorithm that removes a linear trend from the data along the slice and finds the location of the highest emission within each region. This process serves to create a slice which has the background level removed as well as any linear drifts that may occur across the map from mapmaking artifacts such as bowling, which were particularly evident in the JCMT and MUSTANG2 data. Once the slice has been detrended, the signal at the peak intensity that remains is the SED datapoint. 

The method to detrend the data is as follows. First, we apply the slices to the maps. These are defined by their endpoints, between which we draw a line on the map, creating a masked region. Second, at each end point we create a circular mask of radius 20\arcsec in which we calculate the mean value, representing the noise-suppressed value at each end of the slice. Third, we create a linear regression model between the two endpoints of the slice to capture any linear changes across the map as well as the background. Fourth, at each point along the slice we calculate the distance from the endpoint and subtract the linear regression model value. Fifth, we restrict our search to the central 60\% of the slice (targeted on the filament itself) and find the local maximum, this is saved as the location chosen for a given slice and frequency. Once this process has been repeated for all frequencies in a given slice, the SED data locations are extracted by taking the median of the locations chosen in the previous steps to minimize bias introduced by the noise inherent to the data. Once the median locations are calculated, the data are extracted for each frequnecy at that location in the slice and the process is repeated for all slices, producing the initial SEDs used in this analysis. This technique is particularly adept at removing large-scale backgrounds such as the CO contamination in the 90~GHz ACT+Planck maps versus the ACT-only maps, returning datapoints that agree to sub- noise and calibration levels. These data are discussed further in Section \ref{subsec:spectral}.

\begin{figure*}[!htp]
    \centering
    \includegraphics[scale=1]{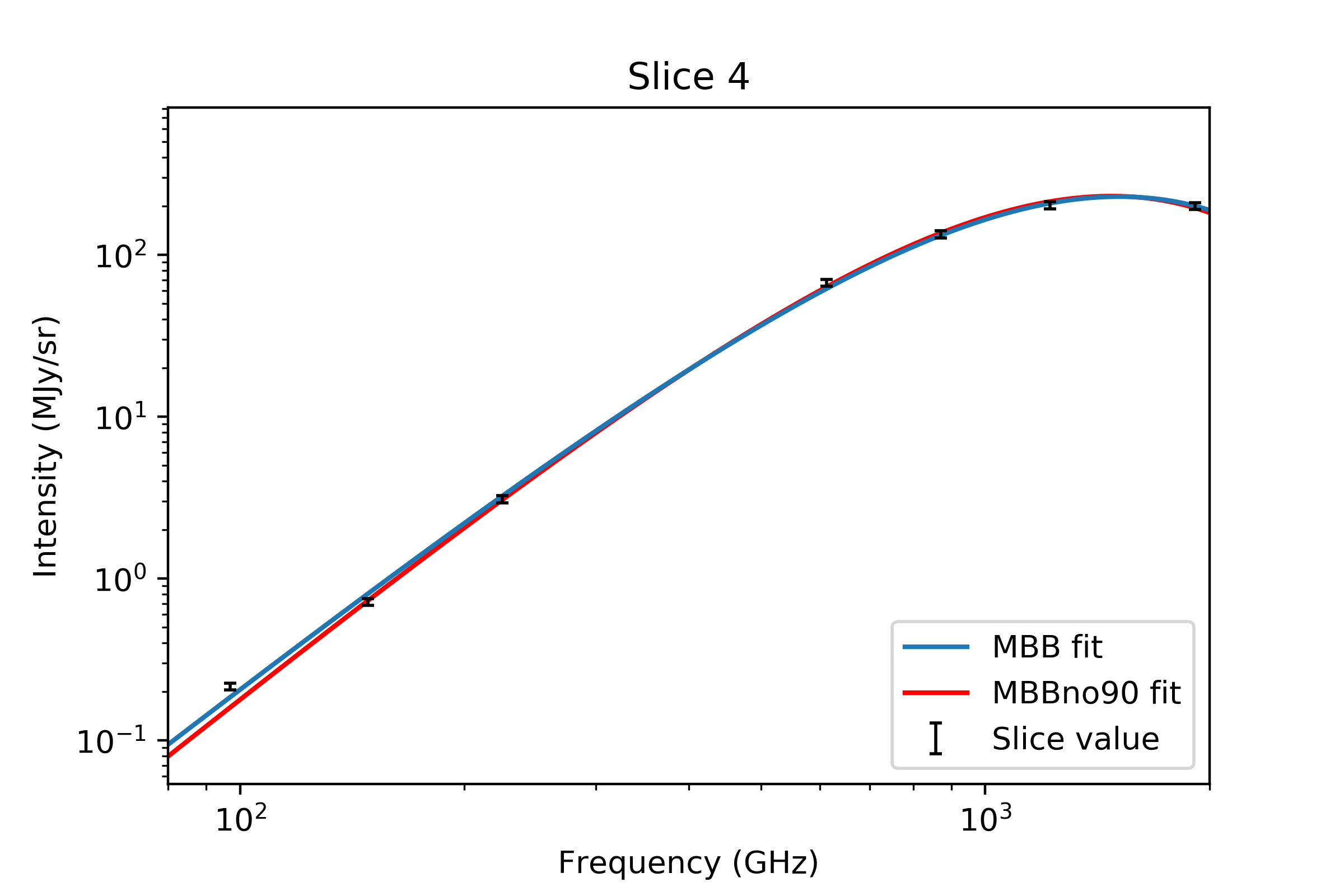}
    \includegraphics[scale=1]{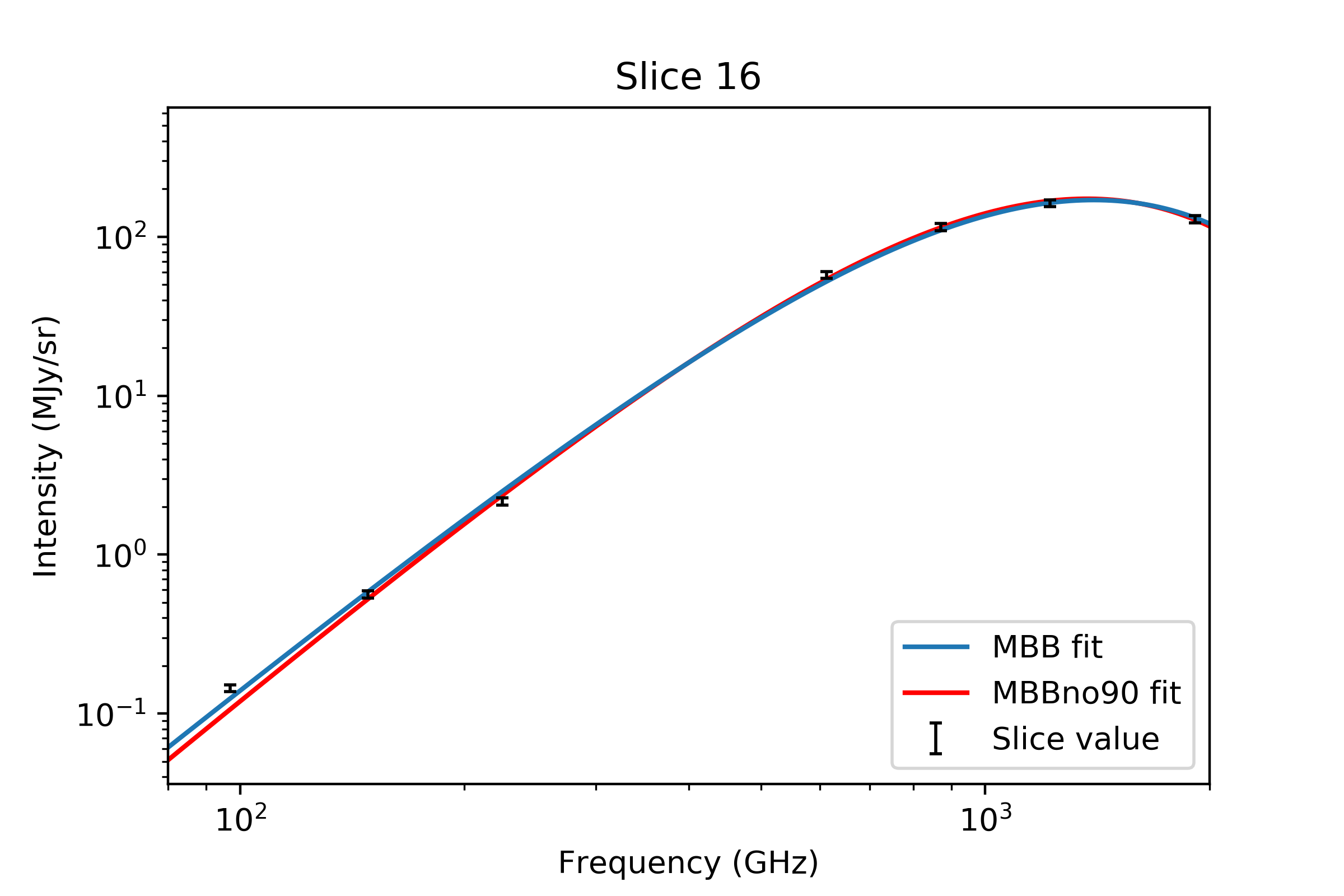}
    \caption{Spectral energy distributions for two example slices in the Orion A-S region at 120\arcsec resolution. The blue lines trace the \textbf{MBBall} and the red lines the \textbf{MBBno90} fit.} 
    \label{fig:OrionA-SEDs}
\end{figure*}

\begin{figure*}[!ht]
    \centering
    \includegraphics[scale=1]{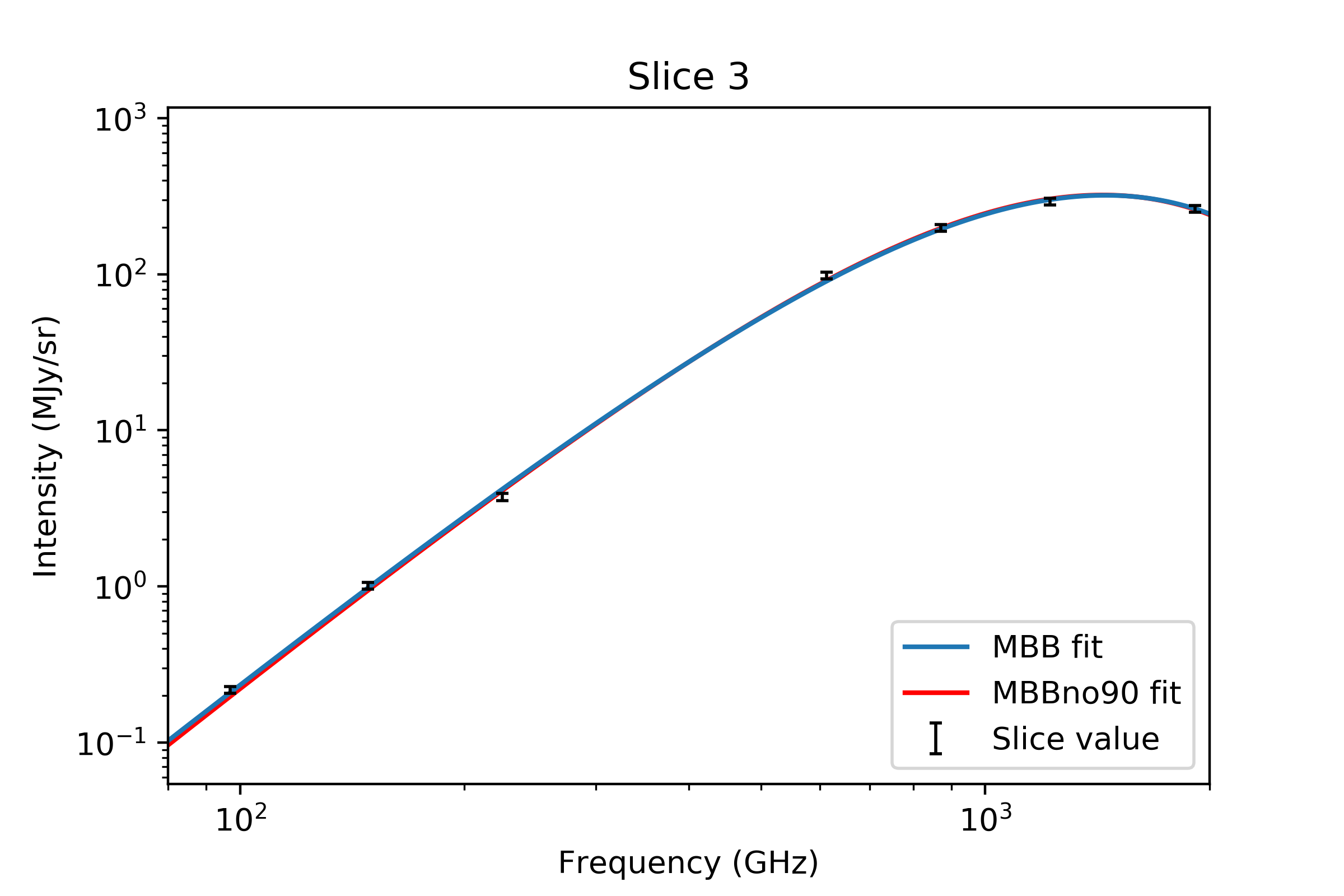}
    \includegraphics[scale=1]{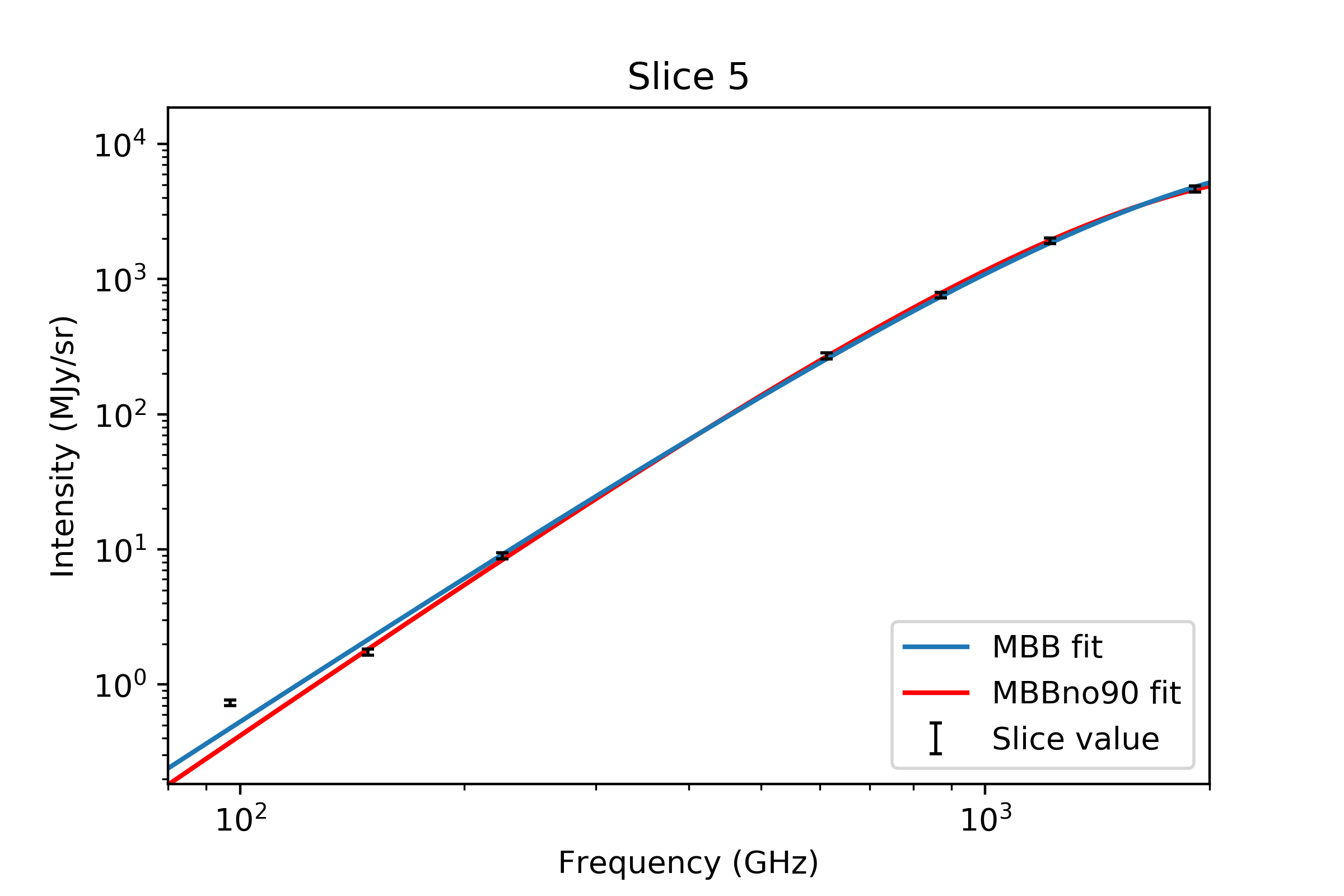}
    \caption{Spectral energy distributions for two example slices in the Orion B region at 120\arcsec resolution. The blue lines trace the \textbf{MBBall} and the red the \textbf{MBBno90} fit.}
    \label{fig:OrionB-SEDs}
\end{figure*}

\begin{figure*}[!ht]
    \centering
    \includegraphics[scale=1]{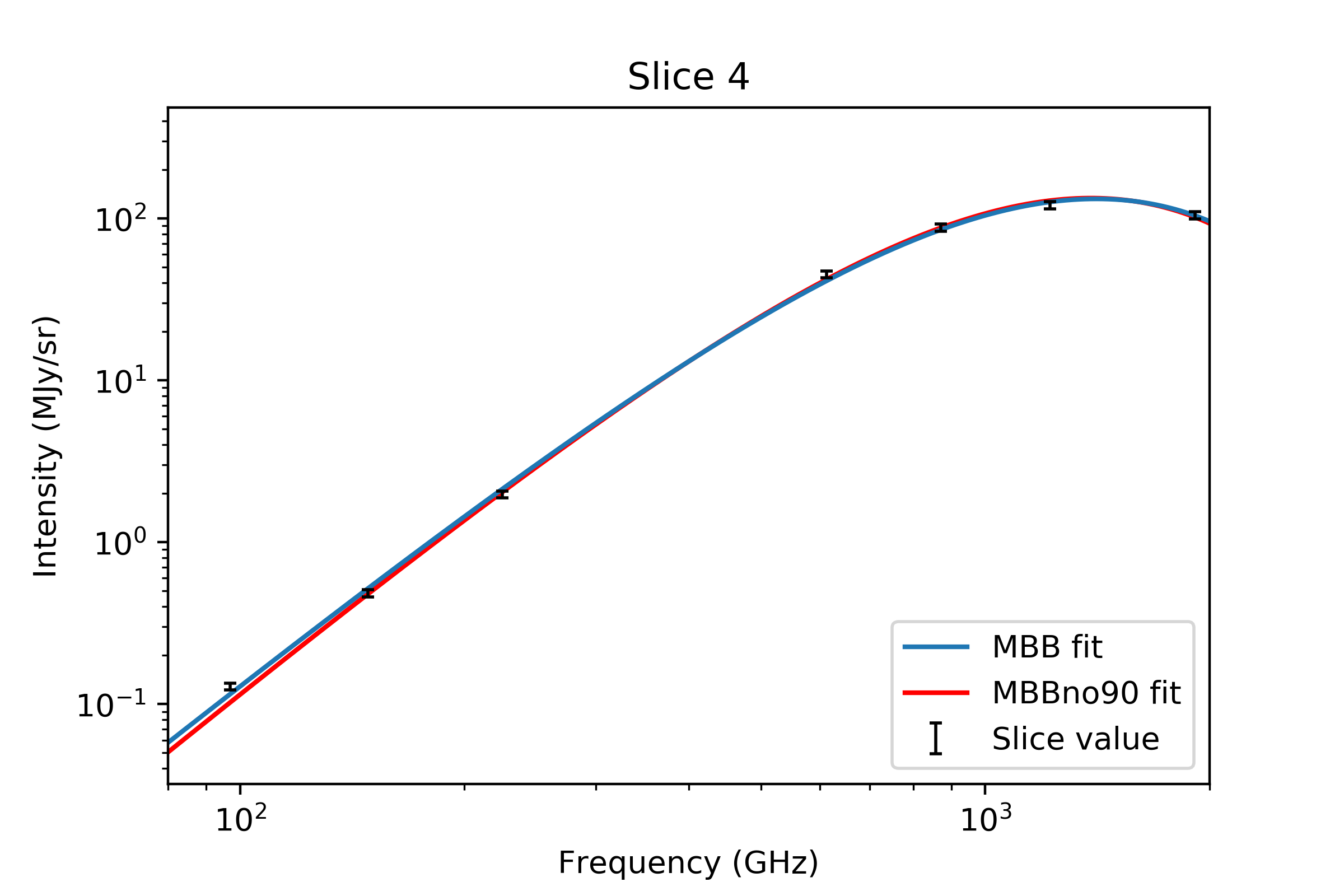}
    \includegraphics[scale=1]{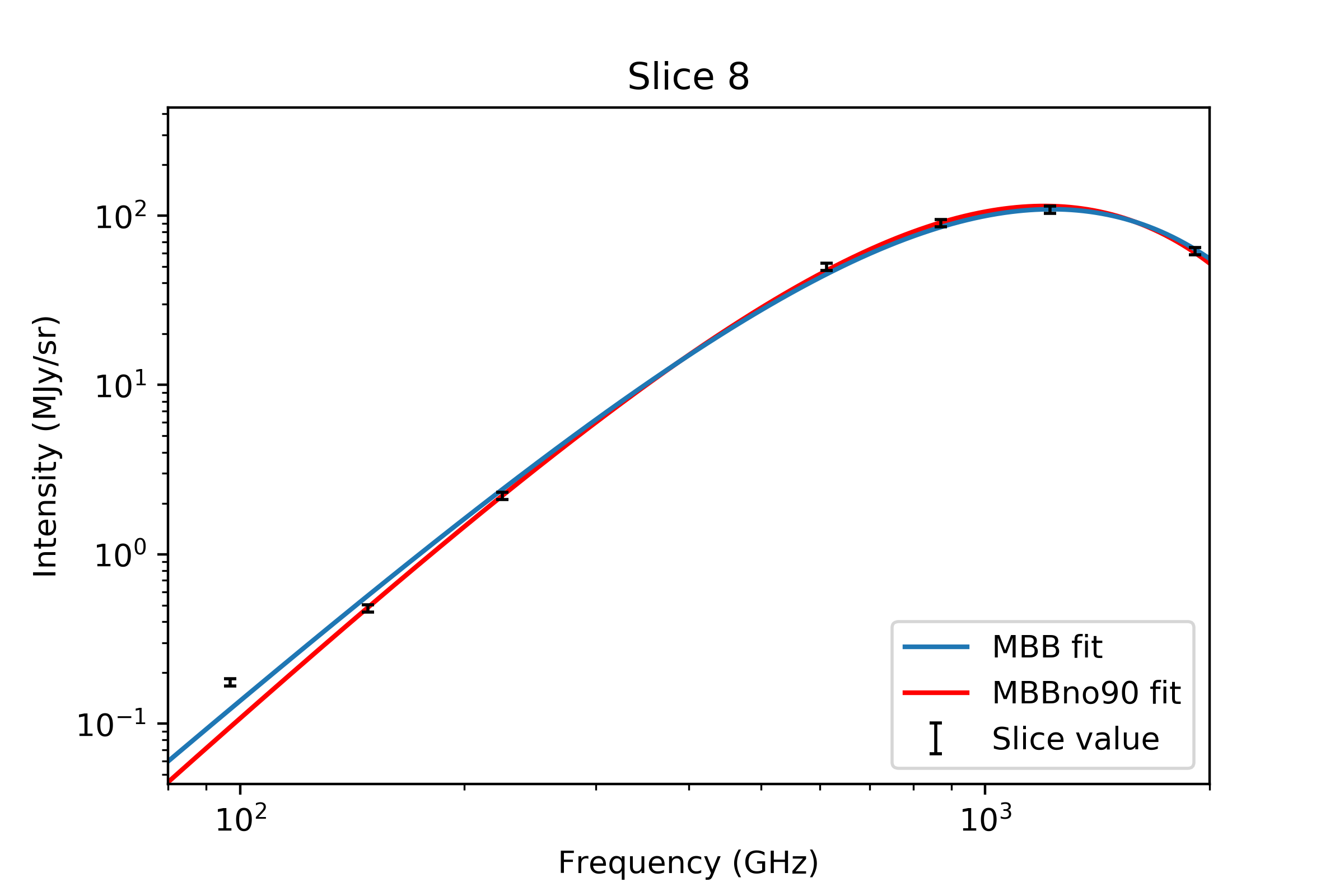}
    \caption{Spectral energy distributions for two example slices in the Serpens 1 region at 120\arcsec resolution. The blue lines trace the \textbf{MBBall} and the red the \textbf{MBBno90} fit.}
    \label{fig:Serpens1}
\end{figure*}

\begin{figure*}[!ht]
    \centering
    \includegraphics[scale=1]{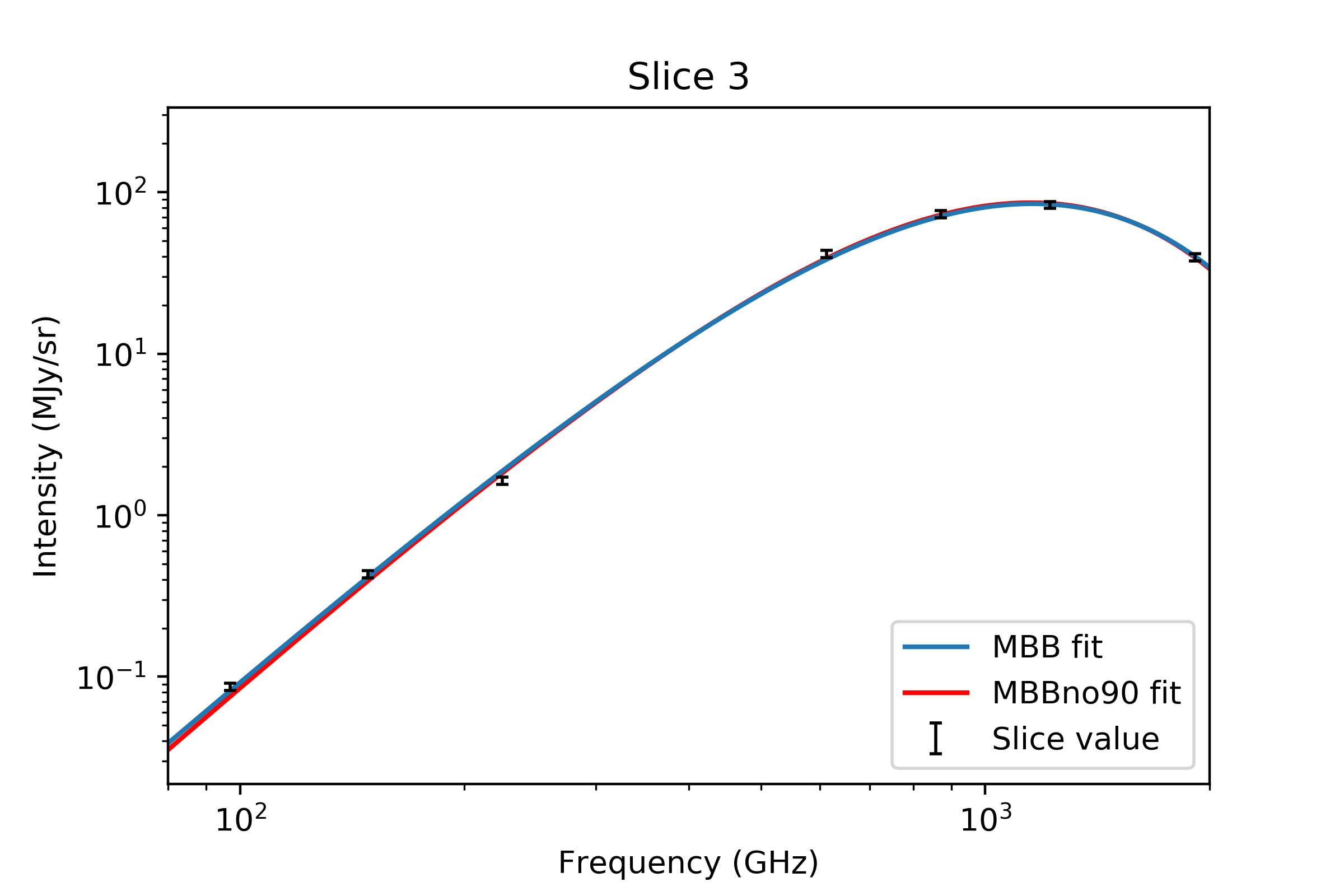}
    \includegraphics[scale=1]{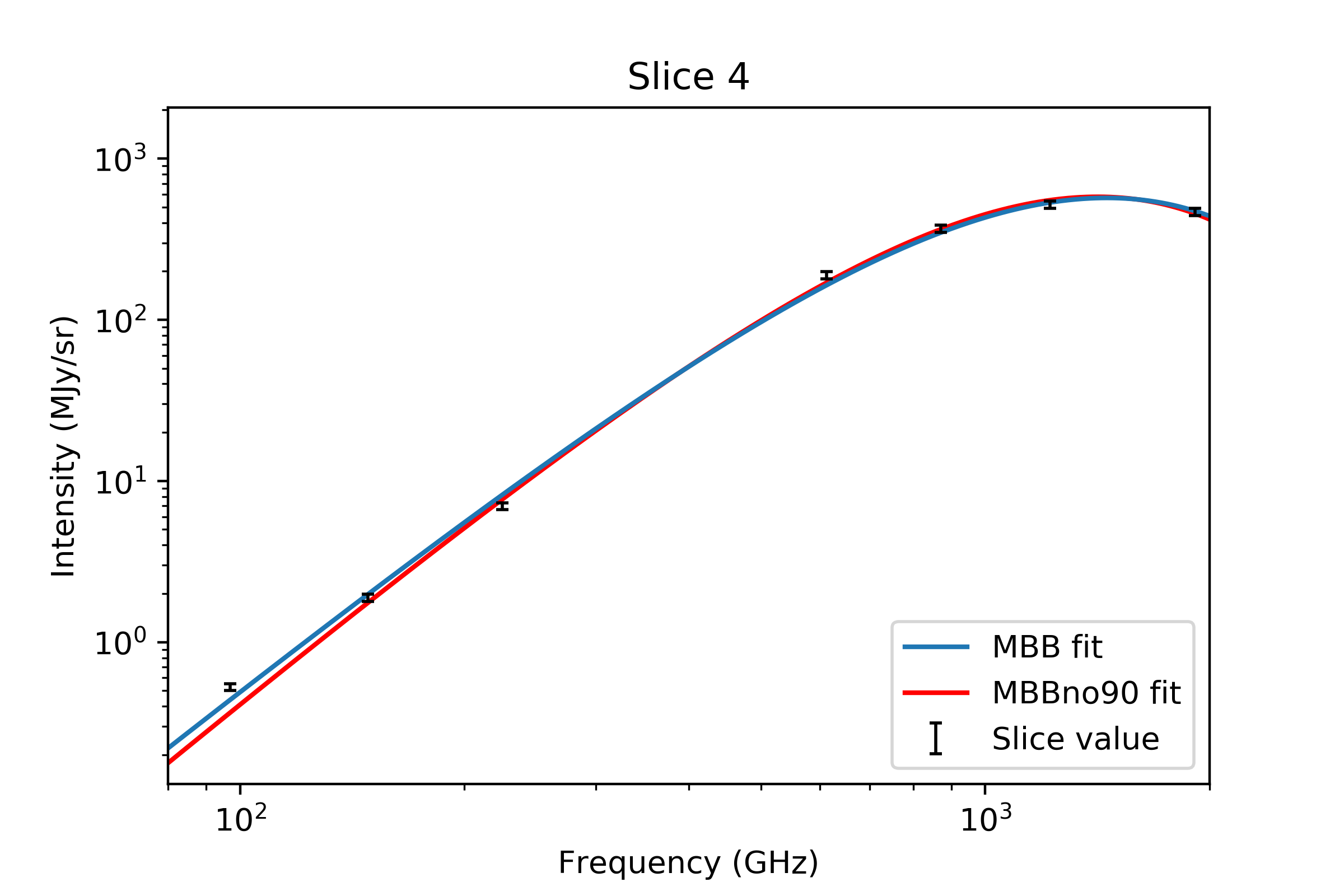}
    \caption{Spectral energy distributions for two example slices slices in the Serpens 2 region at 120\arcsec resolution. The blue lines trace the \textbf{MBBall} and the red the \textbf{MBBno90} fit.}
    \label{fig:Serpens2}
\end{figure*}

\begin{figure*}[!ht]
    \centering
    \includegraphics[scale=1]{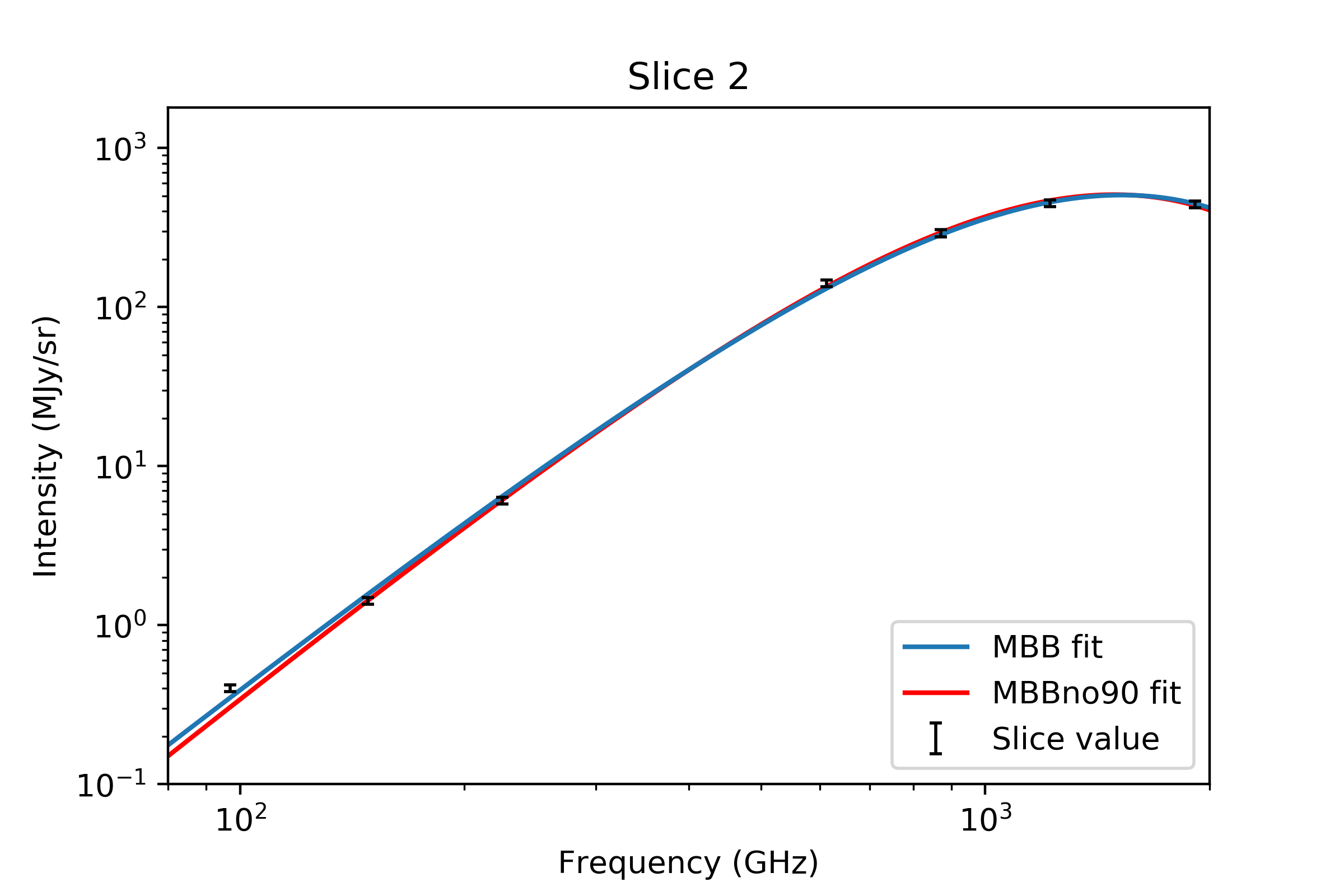}
    \includegraphics[scale=1]{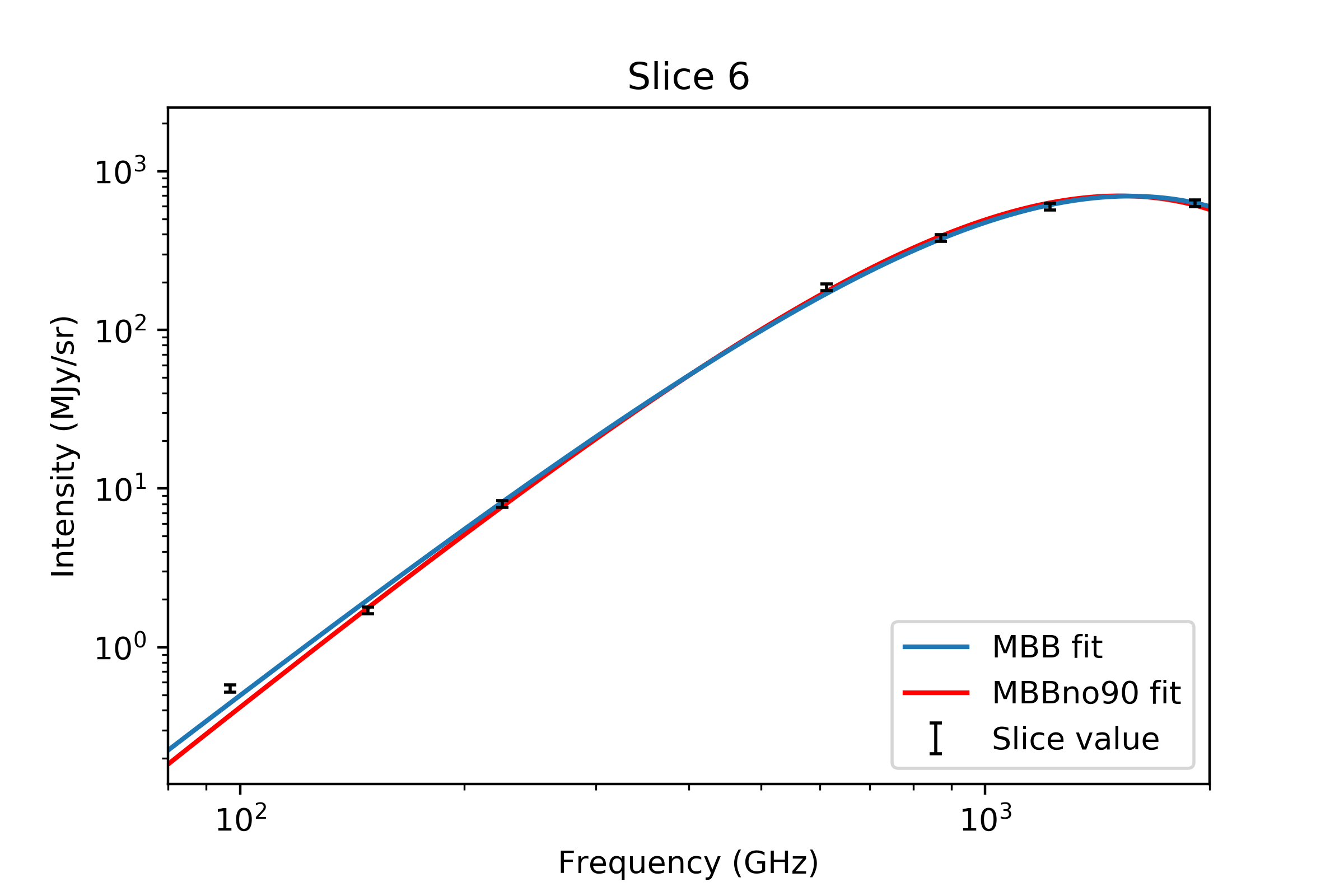}
    \caption{Spectral energy distributions for two example slices in the Serpens 3 region at 120\arcsec resolution. The blue lines trace the \textbf{MBBall} and the red the \textbf{MBBno90} fit.}
    \label{fig:Serpens3}
\end{figure*}

\subsection{SED fitting}
\label{subsec:SED}
\begin{figure}
    \centering
    \includegraphics[scale=.5]{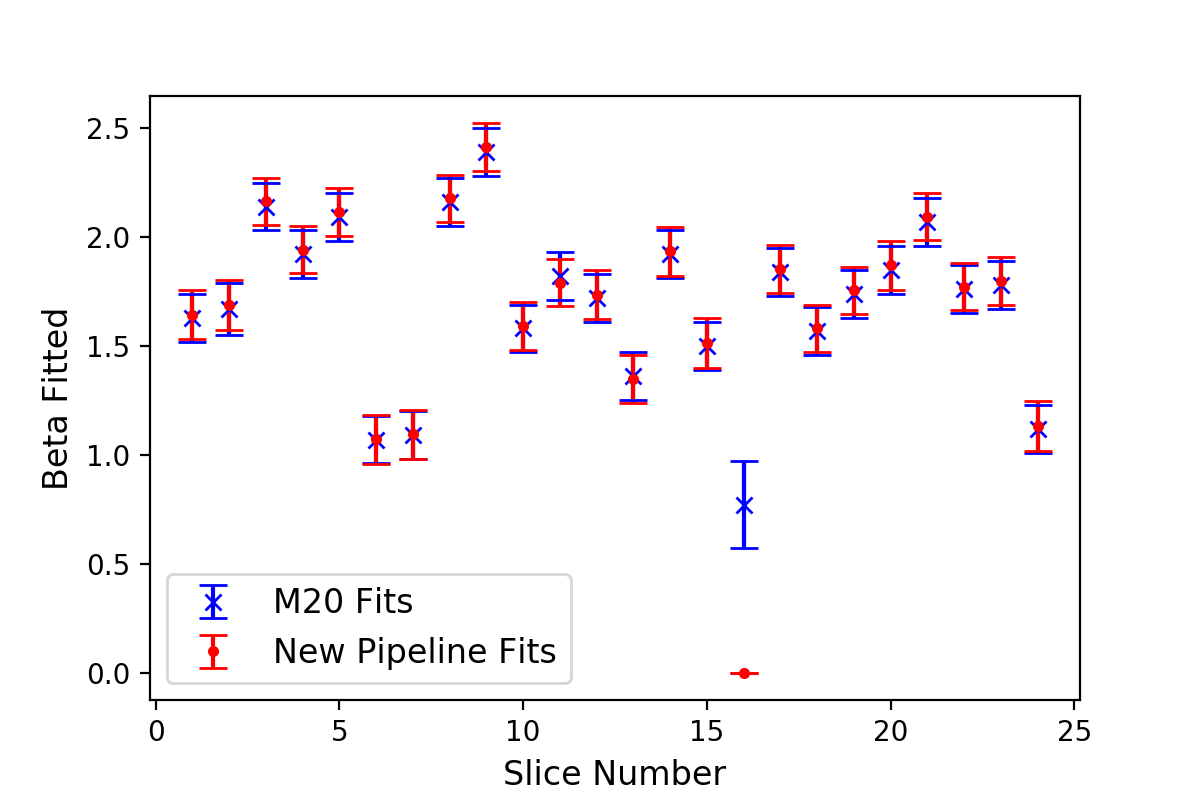}
    \includegraphics[scale=0.5]{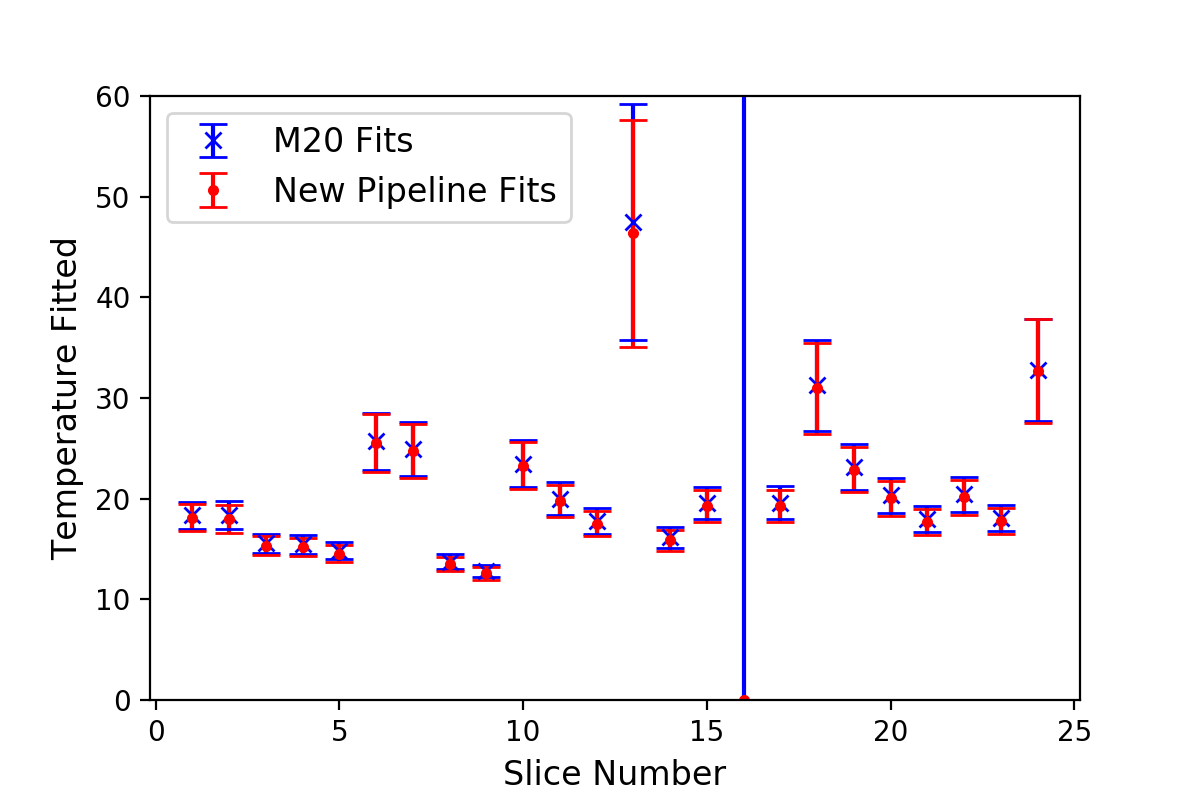}
    \caption{Plots of the calculated parameters $T_d$ and $\beta$ by slice for the M20 and our new pipeline analysis methods for the \textbf{MBBall} fit. Error bars for the red points represent the 1-$\sigma$ uncertainties extracted from the output covariance matrix of the fitting software. For the blue points, the errors are those reported in M20. Of note is that slice 16 comprises poor data which fails to converge to a solution in the new pipeline, and the results of the temperature fit were not reported in the M20 analysis. For these reasons we leave this slice out of the verification analysis.}
    \label{fig:pipecomp}
\end{figure}

With the data extracted from the maps, we apply models to determine the presence of excess emission at low frequencies as well as verify the pipeline consistency with the 25\arcsec data. The dust is modeled to emit as a Modified Blackbody (MBB), which is characterized by free parameters for the dust temperature $T_d$ and spectral index, $\beta$ (see Equation \ref{eq:dust_I}). In order to quantify the level of excess emission at low frequencies, we characterize the behavior of three different fits to the data. They are as follows:
\begin{itemize}
    \item \textbf{Full fit}: We fit the entire SED from 160~$\mu m$ to 3.3~mm with a modified blackbody and characterize the quality of the fit. This is referred to as the \textbf{MBBall} in the SEDs produced.
    \item \textbf{Short $\lambda$}: We fit the SED from 160~$\mu m$ to 2~mm, omitting the 90 GHz point, with a modified blackbody and characterize the quality of the fit to this data. This fit is referred to as the \textbf{MBBno90} fit in the produced SEDs.
    \item \textbf{Short $\lambda$ extrapolated}: Not a truly unique fit, but an extrapolation of the previous down to the 3.3~mm data, allowing us to characterize the quality of that fit to the entire dataset. This serves as a check as to whether the 90~GHz data is significantly different from what the rest of the data would predict it to be. This is referred to as the extrapolated \textbf{MBBno90} fit throughout the text.
\end{itemize}
We fit the data through the use of the \texttt{scipy.optimize} library function \texttt{curve$\_$fit}, which, in combination with the MBB model we provide, performs a least squares fit taking into account the uncertainties in the data provided. The SEDs and both fits for each slice can be found in Figures \ref{fig:omc23-SEDs}-\ref{fig:Serpens3}. We characterize the quality of the fits through the $\chi^2$ per degree of freedom for each variant, the lists of which can be found in Tables \ref{tab:chi-squared-serpens} and \ref{tab:chi-squared-orion}. As discussed in Section \ref{subsec:bandpasses}, we use the response-weighted average frequency $\nu_0$ for the purposes of fitting, residual calculations, and reduced $\chi^2$ values. Verification of the pipeline is discussed in Section \ref{sec:valid} with a comparison of M20 and this analysis fit values found in Figure \ref{fig:pipecomp}.

\subsubsection{SED Monte-Carlo simulations}
\begin{figure}
    \centering
    \includegraphics[scale = 1]{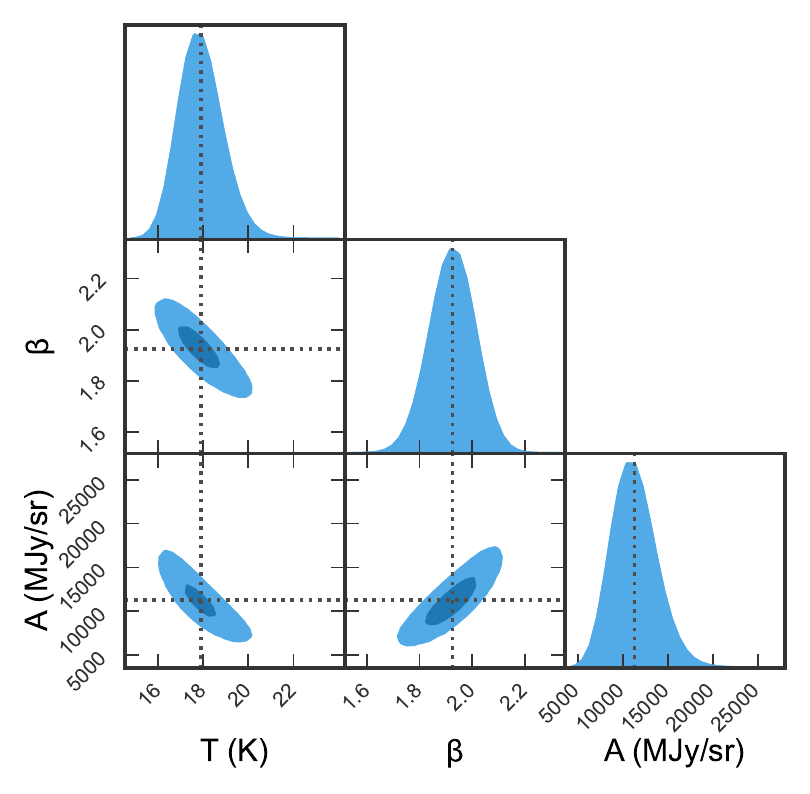}
    \includegraphics[scale = 1]{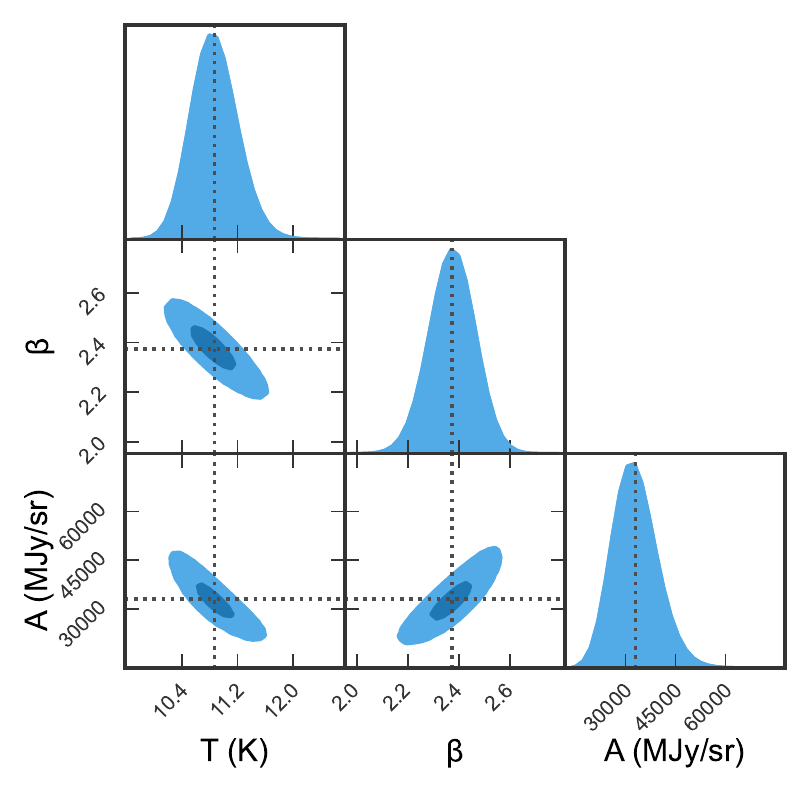}
    \caption{\textbf{A pair of SED parameter corner plots from the MBBno90 fit to the slices. (Top) The SED parameter corner plot for Orion B slice 7, a dataset consistent with no excess emission (see Table \ref{tab:chi-squared-orion}). (Bottom) The SED parameter corner plot for Orion A-S slice 9, which showed $>6\sigma$ excess emission (see Table \ref{tab:chi-squared-orion}). While the SED parameters vary between the two slices, the fitted parameters remain robust in each case.}}
    \label{fig:corner_plot}
\end{figure}

The initial SEDs are constructed from the data that were extracted using the methods described in Section \ref{subsec:extraction}. To quantify the impact of calibration and measurement errors (see Table \ref{tab:errors}), as well as any potential biases inherent in our analysis, we performed a suite of Monte-Carlo simulations. The procedure was as follows. For each slice, the original \textbf{MBBno90} fit values were extracted at the frequencies corresponding to the instruments' bandpass-averaged frequency. Next, a set of $10^6$ Gaussian random values were generated, with a mean value of 1 and a $\sigma_{draw}$ corresponding to the calibration error for each instrument, with the maximally-pessimistic assumption that each instrument (ACT, SPIRE, and PACS) had fully correlated calibration uncertainty. These draws were then applied as multiplicative factors to the original predicted data points and the resulting SED refit, producing a varying suite of parameters (T, $\beta$, and A) as well as new predictions for 90~GHz brightness. \textbf{A pair of corner plots showing the parameter posteriors is displayed in Figure~\ref{fig:corner_plot}, corresponding to a case with no excess emission (top) and a case with a significant 90\,GHz excess (bottom). The parameter posteriors are qualitatively similar in both cases.}

This suite of simulations provided two key pieces of information: an assessment of the bias of the initial SED fits as compared to the noisy realizations and an understanding of the range of predicted values associated with each fit. Notably, for 53 slices and $10^6$ realizations, we find that the initial fits are remarkably unbiased ($\Delta_{90}<0.2\%$, see Tables \ref{tab:chi-squared-serpens} and \ref{tab:chi-squared-orion}), with typical deviations from the final mean parameter values shown in Table \ref{tab:bootstrap}. Finally, we use the distribution of predicted values to characterize the disagreement statistics between the SED predictions and the measured map values, the results of which are discussed in Section \ref{sec:results}.

\subsubsection{90~GHz Monte-Carlo simulations}

In addition to using this procedure to model the effects of high-frequency calibration errors on the SEDs, we also applied this technique to determine the range of possible 90~GHz values. These simulations were constructed in a similar manner - all slices shared the same set of $10^6$ draws for the calibration errors. The key difference comes in adding the map noise values to each slice data point. In this case, the slices were separated into groups corresponding to the particular map they were taken from and the ``dark noise" levels used to create a per-slice additional noise prescription. The resulting collection of data points represent a large sampling of the range of possible values for the 90~GHz emission and providing insight into the significance of the excesses seen in the data.

\begin{table}[]
    \centering
    \begin{tabular}{|c|c|}
    \hline
        Parameter & Mean Deviation (n$_\sigma$) \\ \hline
        Temperature (T) &   0.04        \\ \hline
        Spectral Index ($\beta$) &  -0.007         \\ \hline
        Amplitude at 250$\mu$m (A) & 0.06           \\ \hline
    \end{tabular}
    \caption{The mean deviation of the initial fit parameters across all slices as compared to the $10^6$ Monte-Carlo realizations. As is shown here, the initial fits were slightly biased high on the amplitude and temperature and slightly low on the spectral index. To correct for this, the bootstrap realization values and their associated distributions have been used in lieu of the initial fits throughout this analysis. Of note, these biases reported here would correspond to values of $\Delta T = 0.01K$ and $\Delta \beta = -0.0004$ on average across the entire sample.}
    \label{tab:bootstrap}
\end{table}

\subsubsection{Full data fits (\textbf{MBBall})}
We first attempt a simple modified blackbody fit to cover the entire spectrum of the data, similar to the \textbf{MBBall} performed in the M20 analysis in the OMC 2/3 region. After analyzing the initial OMC 2/3 test region (Figure \ref{fig:omc23-SEDs}, 120\arcsec resolution) which was observed in M20, we looked at the three other regions in Serpens, Orion B, and Orion A-S. In Section \ref{sec:results} we will examine the residuals of this fit more closely region by region.

\begin{table*}
    \centering
    \begin{tabular}{|c|c|c|c|c|c|c|c|}
       \hline
       \multicolumn{8}{|c|}{$\chi^2/n$ (n), \textit{no90} params, excess emission} \\
       \hline
       Location  &  Full fit (n=4) & Short $\lambda$ (n=3) & $\beta$ & $T_d$ & Excess \% &Excess ($\sigma$) & Monte-Carlo ratio\\
       \hline
       \multicolumn{8}{|c|}{\textbf{OMC 2/3}} \\ \hline
       Slice 1 & 1.792&1.071&1.527$\pm$0.083&36.36$\pm$4.78&16.3&2.50 & 1.0019\\ \hline
       Slice 2 & 1.467&1.745&1.669$\pm$0.082&22.28$\pm$1.49&6.2&0.97 & 1.0018\\ \hline
       Slice 3 & 1.002&0.051&1.868$\pm$0.082&17.64$\pm$0.88&16.3& 2.50& 1.0017\\ \hline
       Slice 4 & 0.67&0.087&1.786$\pm$0.082&20.06$\pm$1.18&12.7& 1.97& 1.0018\\ \hline
       Slice 5 & 0.678&0.859&1.884$\pm$0.082&23.25$\pm$1.65&2.9& 0.44& 1.0018\\ \hline
       \multicolumn{8}{|c|}{\textbf{Serpens 1}} \\ \hline
       Slice 1 & 7.346&1.844&1.78$\pm$0.083&14.41$\pm$0.57&47.6&\textbf{6.33} & 1.0018\\ \hline
       Slice 2 & 0.824&0.994&1.92$\pm$0.084&12.28$\pm$0.4&-4.2& -0.72& 1.0018\\ \hline
       Slice 3 & 4.175&1.736&1.906$\pm$0.084&12.09$\pm$0.39&-23.6& -4.01& 1.0018\\ \hline
       Slice 4 & 1.831&1.481&1.847$\pm$0.083&13.83$\pm$0.52&14.0& 2.15& 1.0018\\ \hline
       Slice 5 & 0.45&0.41&1.761$\pm$0.082&17.16$\pm$0.83&-5.7& -0.97& 1.0017\\ \hline
       Slice 6 & 7.956&1.647&1.849$\pm$0.083&13.59$\pm$0.5&50.7&\textbf{6.64} & 1.0018\\ \hline
       Slice 7 & 7.321&0.898&1.996$\pm$0.084&12.17$\pm$0.4&49.5&\textbf{ 6.50}& 1.0018\\ \hline
       Slice 8 & 13.288&0.597&2.114$\pm$0.084&11.27$\pm$0.34&77.7&\textbf{8.87} & 1.0018\\ \hline
       \multicolumn{8}{|c|}{\textbf{Serpens 2}} \\\hline
       Slice 1 & 17.405&0.444&2.062$\pm$0.083&12.42$\pm$0.41&96.4& \textbf{10.06}& 1.0018 \\\hline
       Slice 2 & 3.831&0.692&1.78$\pm$0.082&16.47$\pm$0.76&32.8&\textbf{4.69} & 1.0018 \\\hline
       Slice 3 & 4.341&5.766&2.3$\pm$0.085&10.51$\pm$0.29&2.1&0.30 & 1.0018 \\\hline
       Slice 4 & 7.892&4.018&1.913$\pm$0.083&13.95$\pm$0.53&42.1 & \textbf{5.75}& 1.0018\\\hline
       Slice 5 & 7.939&4.902&1.972$\pm$0.083&12.98$\pm$0.45&38.9& \textbf{5.38}& 1.0018 \\\hline
       \multicolumn{8}{|c|}{\textbf{Serpens 3}} \\\hline
       Slice 1 & 11.824&1.881&1.717$\pm$0.082&15.1$\pm$0.63&67.4 &\textbf{8.14} & 1.0018\\\hline
       Slice 2 & 1.301&0.629&1.838$\pm$0.082&14.92$\pm$0.61&15.0&2.30 & 1.0018 \\\hline
       Slice 3 & 6.192&1.562&1.862$\pm$0.082&14.82$\pm$0.6&42.4& \textbf{5.79}& 1.0018\\\hline
       Slice 4 & 4.087&0.492&1.963$\pm$0.083&13.75$\pm$0.51&35.2& \textbf{4.96}& 1.0018\\\hline
       Slice 5 & 4.193&0.947&2.032$\pm$0.083&12.9$\pm$0.45&33.9 & \textbf{4.79}& 1.0018\\\hline
       Slice 6 & 7.442&1.572&1.864$\pm$0.082&15.06$\pm$0.62&47.4& \textbf{6.32}& 1.0018 \\\hline
       Slice 7 & \textit{\textcolor{red}{11.737}}&\textit{\textcolor{red}{0.581}}&\textit{\textcolor{red}{1.604$\pm$0.082}}&\textit{\textcolor{red}{22.26$\pm$1.49}}&\textit{\textcolor{red}{-}}& \textit{\textcolor{red}{-}}& \textit{\textcolor{red}{-}}\\\hline
    \end{tabular}
    \caption{Reduced $\chi^2$ for the Serpens regions in the two different cases as well as the OMC 2/3 region. The first case is the modified blackbody fit to all data point and the $\chi^2_n$ calculated from these data. The second case is the blackbody spectrum fit to the $\lambda \leq 2mm$ datapoints and the $\chi^2_n$ calculated from the $\lambda \leq 2mm$ dataset as well. The number of degrees of freedom are shown next to each dataset in parentheses. We have additionally included the fit parameters from the bootstrap analysis of the data and the excess at 90 GHz compared to the prediction from the model fit using only the higher frequencies (short $\lambda$ model) is shown as a fractional excess (``Excess \%"). Slices with bolded ``Excess ($\sigma$)" line lie above the 4-$\sigma$ significance level of excess inconsistent with 0. Slice 7 of Serpens 3 was flagged during data quality control due to contamination of the data from a Galactic HII region.}
    \label{tab:chi-squared-serpens}
\end{table*}

\begin{table*}
    \centering
    \begin{tabular}{|c|c|c|c|c|c|c|c|}
       \hline
       \multicolumn{8}{|c|}{$\chi^2/n$ (n), \textit{no90} params, excess emission} \\
       \hline
       Location  &  Full fit (n=4) & Short $\lambda$ (n=3) & $\beta$ & $T_d$ & Excess \% & Excess ($\sigma$) &Monte-Carlo ratio  \\
       \hline
       \multicolumn{8}{|c|}{\textbf{Orion A-S}} \\
       \hline
       Slice 1 & 2.518&1.217&2.123$\pm$0.084&12.02$\pm$0.39&21.9 &3.25 &1.0018\\ \hline
       Slice 2 & 14.715&12.478&2.419$\pm$0.084&11.87$\pm$0.38&47.1& \textbf{6.25}& 1.0018\\ \hline
       Slice 3 & 5.624&4.336&2.165$\pm$0.083&13.12$\pm$0.47&27.5& \textbf{4.00} & 1.0018\\ \hline
       Slice 4 & 5.147&1.136&1.817$\pm$0.082&14.83$\pm$0.60&37.8&\textbf{5.27} & 1.0018\\ \hline
       Slice 5 & 1.358&0.682&2.217$\pm$0.084&11.81$\pm$0.37&15.3& 2.32& 1.0018\\ \hline
       Slice 6 & 4.138&5.167&2.392$\pm$0.084&12.28$\pm$0.40&-7.8& -1.32& 1.0018\\ \hline
       Slice 7 & 6.795&2.301&2.076$\pm$0.084&12.17$\pm$0.40&43.1& \textbf{5.83}& 1.0018\\ \hline
       Slice 8 & 12.259&11.647&2.118$\pm$0.083&12.39$\pm$0.41&36.0&\textbf{5.04} & 1.0018\\ \hline
       Slice 9 & 9.07&3.605&2.373$\pm$0.084&10.87$\pm$0.31&50.2& \textbf{6.55}& 1.0018\\ \hline
       Slice 10 & 9.732&9.227&2.412$\pm$0.084&11.66$\pm$0.36&31.3&\textbf{4.46} & 1.0018\\ \hline
       Slice 11 & 3.169&0.111&1.847$\pm$0.083&13.87$\pm$0.52&31.5& \textbf{4.51}& 1.0018\\ \hline
       Slice 12 & 62.43&15.831&3.065$\pm$0.084&10.84$\pm$0.31&307.0& \textbf{14.78}& 1.0018\\ \hline
       Slice 13 & \textit{\textcolor{red}{52.427}}&\textit{\textcolor{red}{4.528}} &\textit{\textcolor{red}{2.766$\pm$0.084}}&\textit{\textcolor{red}{11.5$\pm$0.35}}&\textit{\textcolor{red}{-}}& \textit{\textcolor{red}{-}}& \textit{\textcolor{red}{-}}\\ \hline
       Slice 14 & 12.349&7.28&2.231$\pm$0.082&16.15$\pm$0.73&52.4& \textbf{6.82}& 1.0018\\ \hline
       Slice 15 & 17.628&5.114&2.563$\pm$0.084&11.58$\pm$0.36&84.1&\textbf{9.31} & 1.0018\\ \hline
       Slice 16 & 6.411&2.94&2.029$\pm$0.083&13.14$\pm$0.47&38.6& \textbf{5.35}& 1.0018\\ \hline
       Slice 17 & 2.513&2.108&2.396$\pm$0.084&10.89$\pm$0.32&16.4&2.47 & 1.0018\\ \hline
       \multicolumn{8}{|c|}{\textbf{Orion B}} \\ \hline
       Slice 1 & 0.242&0.302&1.755$\pm$0.082&18.75$\pm$1.01&-1.9& -0.34& 1.0017\\ \hline
       Slice 2 & 7.429&7.594&1.807$\pm$0.083&13.57$\pm$0.5&23.9&3.53 & 1.0018\\ \hline
       Slice 3 & 2.689&2.947&1.935$\pm$0.083&13.86$\pm$0.52&11.5& 1.78& 1.0018\\ \hline
       Slice 4 & 28.362&0.17&1.687$\pm$0.082&28.65$\pm$2.68&146.7&\textbf{12.21} & 1.0018\\ \hline
       Slice 5 & 14.836&1.734&1.884$\pm$0.082&26.37$\pm$2.21&79.0& \textbf{9.03}& 1.0018\\ \hline
       Slice 6 & 18.27&7.654&2.158$\pm$0.082&16.82$\pm$0.79&80.3&\textbf{ 9.10}& 1.0018\\ \hline
       Slice 7 & 41.204&6.021&2.3$\pm$0.083&14.03$\pm$0.54&199.9& \textbf{13.47}& 1.0018\\ \hline
       Slice 8 & 0.400&0.523&1.925$\pm$0.082&17.9$\pm$0.91&-1.3&-0.24 & 1.0017\\ \hline
       Slice 9 & 1.193&0.736&1.801$\pm$0.082&23.92$\pm$1.76&13.1& 2.02& 1.0018\\ \hline
       Slice 10 & 2.062&0.215&1.892$\pm$0.082&16.11$\pm$0.72&23.8& 3.53& 1.0018\\ \hline
       Slice 11 & 6.485&1.26&1.779$\pm$0.082&19.34$\pm$1.08&43.7& \textbf{5.95}& 1.0017\\ \hline
    \end{tabular}
    \caption{Reduced $\chi^2$ for the OrionA-S and OrionB regions in the two different cases. The first case is the modified blackbody fit to all data point and the $\chi^2_n$ calculated from these data. The second case is the blackbody spectrum fit to the $\lambda \leq 2mm$ datapoints and the $\chi^2_n$ calculated from the $\lambda \leq 2mm$ dataset as well. The number of degrees of freedom are shown next to each dataset in parentheses. We have additionally included the fit parameters from the bootstrap analysis of the data and the excess at 90 GHz compared to the prediction from the model fit using only the higher frequencies (short $\lambda$ model) is shown as a fractional excess (``Excess \%"). Slices with bolded ``Excess ($\sigma$)" line lie above the 4-$\sigma$ significance level of excess inconsistent with 0. Note that slice 13 for OrionA was added as a "sanity check" and includes a strong 90 and 150 GHz source that breaks the blackbody curve. It was not included in any calculations due to the known contamination.}
    \label{tab:chi-squared-orion}
\end{table*}

\subsubsection{Short wavelength $\lambda \leq 2~mm$ fits (\textbf{MBBno90} fit)}
With the entire spectrum modeled, we follow the methods of M20 and consider how a spectrum omitting the 3~mm point fits the data. We first restrict ourselves to a MBB fit which covers the range from 160~$\mu m$ to $2~mm$ and consider the quality of this fit to the data. The results of this model are covered in-depth in Section \ref{sec:results}. Restricting the data to higher frequencies improves the quality of the fits, which are shown as the green curves in Figures \ref{fig:omc23-SEDs}-\ref{fig:Serpens3}. If the emission is elevated as we seem to see, we should expect a trend of increased deviation from the model at 3~mm relative to that seen with the \textit{full fit}, as well as an increased $\chi^2_n$ as compared to both other models. Not only do the 90~GHz data lie above the modeled SED curves, but when the fits exclude the 90 GHz data, there is a systematic and large increase in the amount by which the 90~GHz points lie above the curves, clearly showing an excess emission level compared to expectations. As with the previous analyses, the results of this are presented region by region in Section \ref{sec:results}.

\subsection{Further modeling}
\label{sec:furthermodels}
We also consider further approaches to fitting these spectra from a ``quality of fit" perspective. The first approach is to consider a two-component dust population which models the SED as a combination of two components at different temperatures that emit with different spectral characteristics. We follow \citet{Meisner2015}, hereafter M15, and consider the model
\begin{equation}
\begin{aligned}
    I_{\nu, Dust} = A(\frac{\nu}{\nu_0})^{3+\beta_1}\frac{1}{exp(h\nu/kT_{d,1})-1} \\
    +B(\frac{\nu}{\nu_0})^{3+\beta_2}\frac{1}{exp(h\nu/kT_{d,2})-1}.
\end{aligned}
\end{equation}
In this model we have wrapped together the fraction and optical efficiency terms \textit{$f_n
$} and \textit{$q_n$} as well as the overall proportionality constant into a pair of terms A and B which function as the amplitudes of the two dust populations. With our relatively limited number of data points, we opt to use the best-fit parameters found in M15 for $\beta_1$ = 1.67 and $T_1$ = 9.15 to model a second, cool component of the dust emitting in tandem. We then fit the amplitude of this cool component and a full, second modified blackbody to the SEDs. The results of these fits are discussed in Section \ref{subsec:2comp}.

The second model we consider is that of \citet{Paradis_2011}, in which the grains are treated as an amorphous solid, with an emissivity that differs from the standard treatment. This SED is modeled as 
\begin{equation}
    I_{\nu} \propto \epsilon(\nu, T_0)\frac{\nu^3}{exp(h\nu/kT_d)-1}.
\end{equation}
Where they have calculated the greybody $\epsilon$ term as a function of frequencies and temperatures for a variety of datasets which cover the spectrum that we are modeling within. As was tested in M20, we find that the $T_0$ = 17.5K model provides a reasonable estimate of the temperatures of the dust we model ($\sim$15K average temperature) and use the opacities that correspond to that dust temperature as that measured constant is the closest fit to our data. That said, we do not constrain the model to assume a 17.5~K dust temperature, but allow that to be fit as well. The results of this model are discussed in Section \ref{subsec:amorph}.

\section{Validation and concerns}
\label{sec:valid}

\subsection{Verifcation of pipeline at 25\arcsec}
The fitting method that is applied here, as well as the data extraction methods are slightly different from those found in M20, and as such we expect a slight variance in the MBB parameters for identical slices at 25\arcsec. In order to verify that our pipeline produces reasonable results, we fit the data for the 24 slices extracted with our pipeline using the datasets at 25\arcsec from M20. The resulting parameters are compared with those from M20 to ensure that the two methods are in good agreement. A graphical comparison of the fitted parameters with their associated errors can be seen in Figure \ref{fig:pipecomp}. When we compare the solutions to the least-squares fit that each algorithm finds, we see that there exists a small, but consistent difference in the fit parameters across the majority of the slices. In the case of the fit temperatures $T_d$ in each slice, we find that our pipeline predicts a higher temperature across all slices, with a median difference and error of $\Delta T_d/\sigma_{T_d}$ = -0.190$\pm 0.095$. For the spectral indices, we find that this trend is reversed, with our pipeline solving for a slightly lower spectral index across most slices, for median difference and error of $\Delta \beta/\sigma_{\beta}$ = 0.13$\pm 0.11$. We accept these small variances as reasonable and associate them with the well-known degeneracy between $\beta$ and T \citep{Shetty_2009}, the different implementations of the slicing and deramping method, and the least-squares fitting algorithms.

\subsection{Free-Free considerations}
Of the most concern for contamination of the SEDs at 90~GHz is  free-free emission, which has a $\nu^{-0.1}$ spectrum. In both S16 and M20, this was treated through the examination of serendipitous VLA X-band data which overlapped with the OMC 2/3 region.  Of the 24 slices examined in M20, only one was found to have any non-negligible flux associated with free-free emission. To ameliorate concerns that this contamination might provide false positives for enhanced emission in our maps, we used the NRAO VLA Sky Survey (NVSS) catalog and postage-stamp database \citep{Condon_1998} to identify objects which were potential sources of contamination near or on the data taken from each slice. This was then cross-referenced with the Wide-field Infrared Survey Explorer (WISE) catalog of HII regions \citep{Anderson_2014}. This allowed me to determine if the sources were simply 1.4~GHz with a quickly falling off (synchrotron) or slowly dropping (free-free) spectrum. Of the sources identified as potentially problematic (pixel flux assuming free-free spectrum $\geq5\%$ of the flux at 90~GHz) only slice 7 of the Serpens 3 region was associated with an HII region (G029.825+02.232) within a distance of 15\arcmin. Of note, Slice 13 of the Orion A-S region was also flagged and coincided with an HII region; however, that data was already excised as the ACT 90 and 150 maps had shown a strong source not seen in other frequencies and was originally included solely to test the code.  We flagged the Serpens 3 Slice 7 region as contaminated at a $>5\sigma$ level and removed it from the calculations.

\subsection{Spectral line contamination}
\label{subsec:spectral}
\begin{figure}
    \centering
    \includegraphics[scale=0.35]{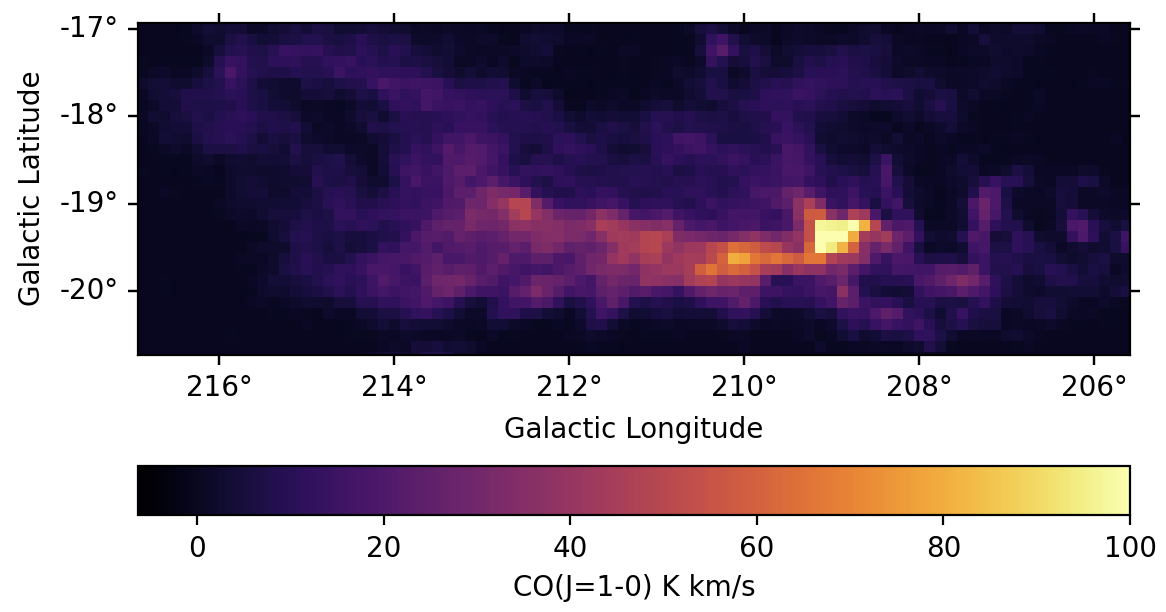}
    \caption{Measured CO(J=1-0) emission in the Orion A molecular cloud complex from \citet{dame_co_orion}. This analysis avoids the bright HII region in between seen on the right side of the image, and is focused on the trailing tail, where the emission levels are between 50-80 K km/s.}
    \label{fig:COmap}
\end{figure}

In addition to the free-free considerations discussed above, we considered the potential for map contamination through molecular emission lines. Of particular interest are the coupling of the 115~GHz CO(J=1-0) line to the \textit{Planck} 100 GHz maps which are used in the production of the ACT dataset and the HCN emission line at 88.6~GHz which sits within all bandpasses of the ACT 90~GHz maps. To evaluate this systematic concern, we reproduced the analysis with the ACT-only maps, which are not sensitive to the CO transition as it lies outside the bandpass, and found that the results agreed with those of the ACT+Planck maps to within a maximum discrepancy of 3\%. The background subtraction tests in the Orion A-S area corresponded to values of 60-80 K km/s CO(J=1-0) brightness, in good agreement with the survey data shown in Figure \ref{fig:COmap} with typical values of 50-80 K km/s \citep{dame_co_orion}. Our procedure is robust to CO contamination in the Planck passband because of the size of the \textit{Planck} beam (9.65\arcmin, \citet{planck2014_7}), which is the same size scale as the slices and the CO emission turns up overwhelmingly as a constant background level. This background is removed during the slicing and deramping process. 

The remaining problematic line is the HCN line, which has potential for contamination in all instruments used in these analyses. To estimate the level of contamination from this line, we scaled the conversion factor found in \citet{planck2014_13} for CO by the bandpass ratio and the emission frequencies for HCN vs CO and compared the expected signal from HCN in these regions to the excess seen in the 90~GHz maps. Estimates of the typical emission levels were found in \citet{1981_HCN} and \citet{OrionA-HCN} which showed peaks in the densest regions of $\sim 60$ K km/s and more typical levels throughout the cloud of $\sim$5-6. In order to account for the excess signal seen at 90~GHz, all regions would need to be in excess of 65 K km/s HCN brightness, which is more than 10x the typical levels seen across the region. In sum, we did not find any evidence that CO or HCN emission can account for the excess signal level seen in this work.
\section{Interpretation}
\label{sec:results}

\subsection{OMC 2/3}

The first region we consider in detail is the OMC 2/3 region that was examined in M20. For the \textbf{MBBall} fit model and 120\arcsec  resolution, we find that when we apply this method to the 5 slices in the OMC 2/3 region we see an excellent agreement with the model with a median $\chi^2_n=\chi^2/n$ = 1.00 for n=4 degrees of freedom. In this region the minimum was $\chi^2_n$ = 0.66 and the maximum was $\chi^2_n$ = 1.79. As did M20, we examined the 3~mm datapoints and find that, while they lie above the SED in all cases, they are much more in line with the model with an average excess of 4.9\%. When examined at 25\arcsec we saw a mean brightness excess of 32$\%$, a far more noticeable discrepancy than at 120\arcsec. This smaller difference between 90~GHz and the model contributes to the significantly better agreement and fit seen in our analysis.

We next consider the results of the \textbf{MBBno90} fit, and find that the median was $\chi^2_n$ = 0.85 for n=3 degrees of freedom, the minimum was $\chi^2_n$ = 0.05, and the maximum was $\chi^2_n$ = 1.74 across the five slices in this map. The drop in the median relative to the full data fit comes as a result of slices 3 and 4, both of which are nearly perfectly fit in this regime by the model. The final model is the \textbf{MBBno90} fit extrapolated down to 90~GHz which gives us a median $\chi^2_n$ = 1.65 for n=4 degrees of freedom, a minimum of $\chi^2_n$ = 0.72, and a maximum of $\chi^2_n$ = 2.77 across 5 slices with a notable increase in all of the slices. Looking at the deviation of the 90~GHz datapoint we see a mean excess of 10.9$\%$. It appears that there is slight evidence that the 90~GHz data is elevated with respect to the model when it is constrained by that data, and that the evidence becomes significant when the model is not constrained by that data. Unfortunately, in this region, none of excesses seen broke the 4-$\sigma$ threshold we have set, though 2/5 were above a 2-$\sigma$ level. Even though less excess is seen at 120\arcsec as compared to the large amount seen at 25\arcsec, we do not speculate as to the cause of this as the conclusions come from disparate datasets, the slices do not coincide with each other, and we lack additional data for comparison in other regions.

\subsection{Serpens}

The Serpens region from the HGBS contains a total of 20 slices, of which 19 are uncontaminated (Section \ref{subsec:spectral}), split into three sub regions for data processing. Using the \textbf{MBBall} fit method and combining the data into one set, we obtain a median $\chi^2_n$ = 6.75 with a minimum of $\chi^2_n$ = 0.45 and a maximum of $\chi^2_n$ = 17.40. We attribute this degraded consistency with the MBB to the fact that in  17/19 slices, we find that the 3~mm data point lies above the SED curve, with an average excess of 17.1\%. When we apply the \textbf{MBBno90} fit to this region, we find a median $\chi^2_n$ = 1.23 for n=3 degrees of freedom, a minimum of $\chi^2_n$ = 0.41, and a maximum of $\chi^2_n$ = 5.76. In this case, we see a reduction in $\chi^2_n$ in 17/19 slices, which tracks with the excess emission seen at 90~GHz in 17/19 slices. We then applied the extrapolation to 90~GHz and found that a median $\chi^2_n$ = 11.17 for n=4 degrees of freedom, a minimum of $\chi^2_n$ = 0.67, and a maximum of $\chi^2_n$ = 24.42 with an increase compared to both other models visible in all slices. For this region, we also see an average excess emission to 34.7\% relative to the model. Additionally, we found that for 13/19 slices, the excess at 90~GHz is significant to at least a 4-$\sigma$ level.

\subsection{Orion B}

In the Orion B region we measured a total of 11 slices. The \textbf{MBBall} fit model applied to the data returned a median $\chi^2_n$ = 6.48 with a minimum of $\chi^2_n$ = 0.24 and a maximum of $\chi^2_n$ = 41.20. As with the Serpens region, we find that this reduced consistency with the model as compared to the OMC 2/3 region stems from the fact that most (9/11) of the 3~mm measurements are far above the SED, displaying a mean excess of 30.7\%. The results of the \textbf{MBBno90} fit are in line with the other regions', showing a drop to a median $\chi^2_n$ = 1.26 for n=3 degrees of freedom with a minimum of $\chi^2_n$ = 0.17 and a maximum of $\chi^2_n$ = 7.65. We also see that again, most slices (7/11) have a reduction in the $\chi^2_n$ value while the others remain nearly constant, which we expect given the excess emission seen in 9/11 slices. When extrapolated down to 90~GHz, the fit degrades to a median $\chi^2_n$ = 9.41 for n=4 degrees of freedom, a minimum of $\chi^2_n$ = 0.26, and a maximum of $\chi^2_n$ = 48.90. Additionally, the mean excess seen in the 90~GHz data points within Orion-B is increased to 56.2\% and for 5/11 slices the excess is significant to at least a 4-$\sigma$ level.

\subsection{Orion A-S}

The final region is the Orion A-S region which contains a total of 17 slices, of which 16 are not contaminated by an HII region (Section \ref{subsec:spectral}). We applied the \textbf{MBBall} fit to this region and found a median $\chi^2_n$ = 6.79 with a minimum of $\chi^2_n$ = 1.36 and a maximum of $\chi^2_n$ = 62.43. Again, we find that the reduced consistency with an MBB model is expected, as 15/16 of the slices show an elevated emission level at 3~mm, with an average excess of 27.1\%. Applying the \textbf{MBBno90} fit next, we see a median $\chi^2_n$ = 3.97 for n=3 degrees of freedom, a minimum of $\chi^2_n$ = 0.11, and a maximum of $\chi^2_n$ = 15.83. As with the rest of the regions, we see a majority of slices (13/16) with a reduction in the $\chi^2_n$ value as the model is able to better fit the short wavelength data. With these fits examined, we looked at the extrapolation of the \textbf{MBBno90} fit to 90~GHz and found that this new fit had a median $\chi^2_n$ = 10.79 for n=4 degrees of freedom, a minimum of $\chi^2_n$ = 2.27, and a maximum of $\chi^2_n$ = 68.78 with an increase versus both models in 14/16 slices. Additionally, the average elevation of the 90~GHz data point had increased to 52.0\% with a 4-$\sigma$ or higher significance in 12/16 slices.

\begin{figure}
    \centering
    \includegraphics[scale=0.55]{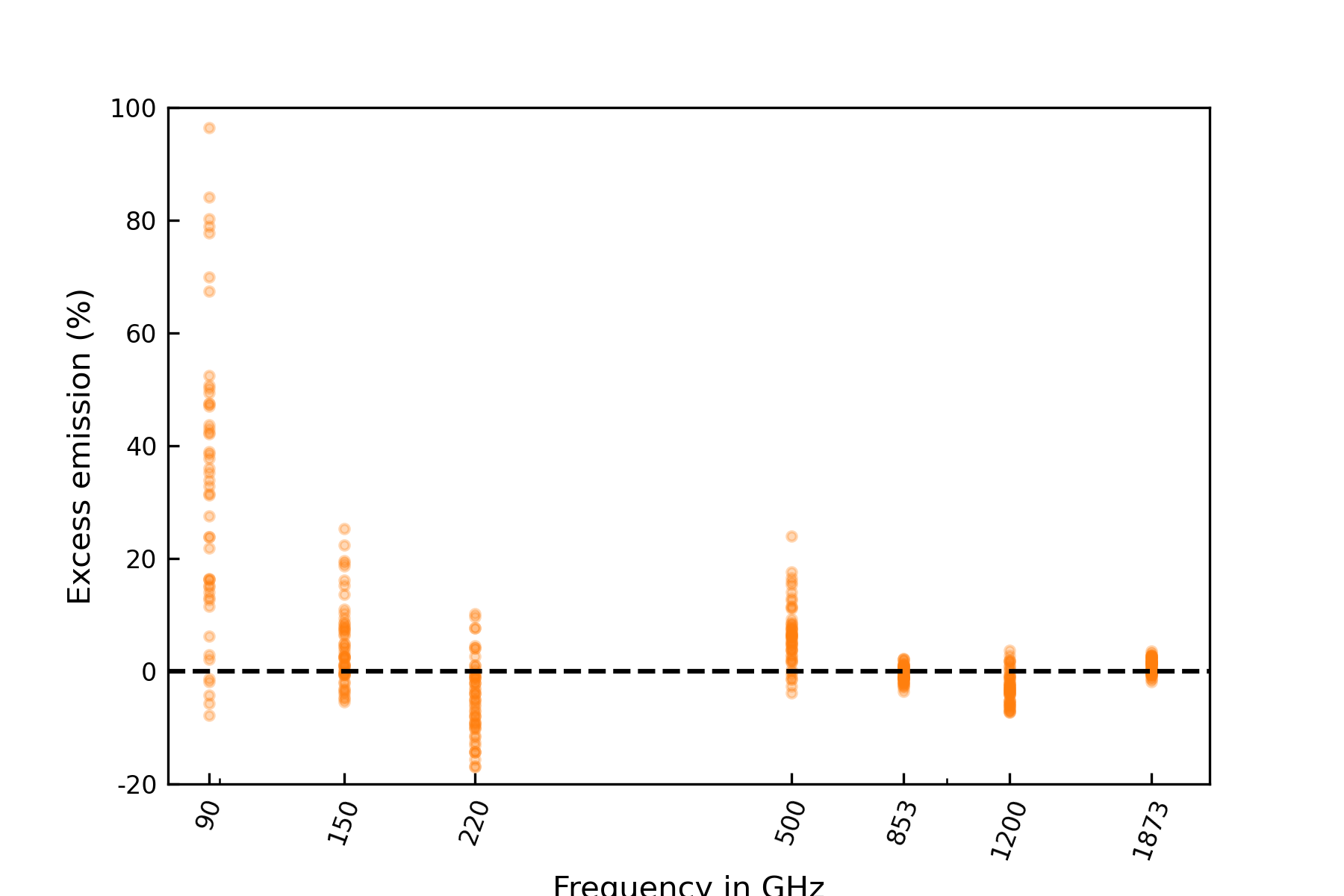}
    \caption{The residuals of the \textbf{MBBno90} fit in units of the \%-excess flux relative to the model as a function of frequency. Only the 90 GHz point is a significant deviation from the model as seen in the scatter near 0 for the 150~GHz to 1.8~THz points and a dramatic rise to over 40\% at 90~GHz.}
    \label{fig:meanexcesses}
\end{figure}

\begin{figure}
    \centering
    \includegraphics[scale=0.5]{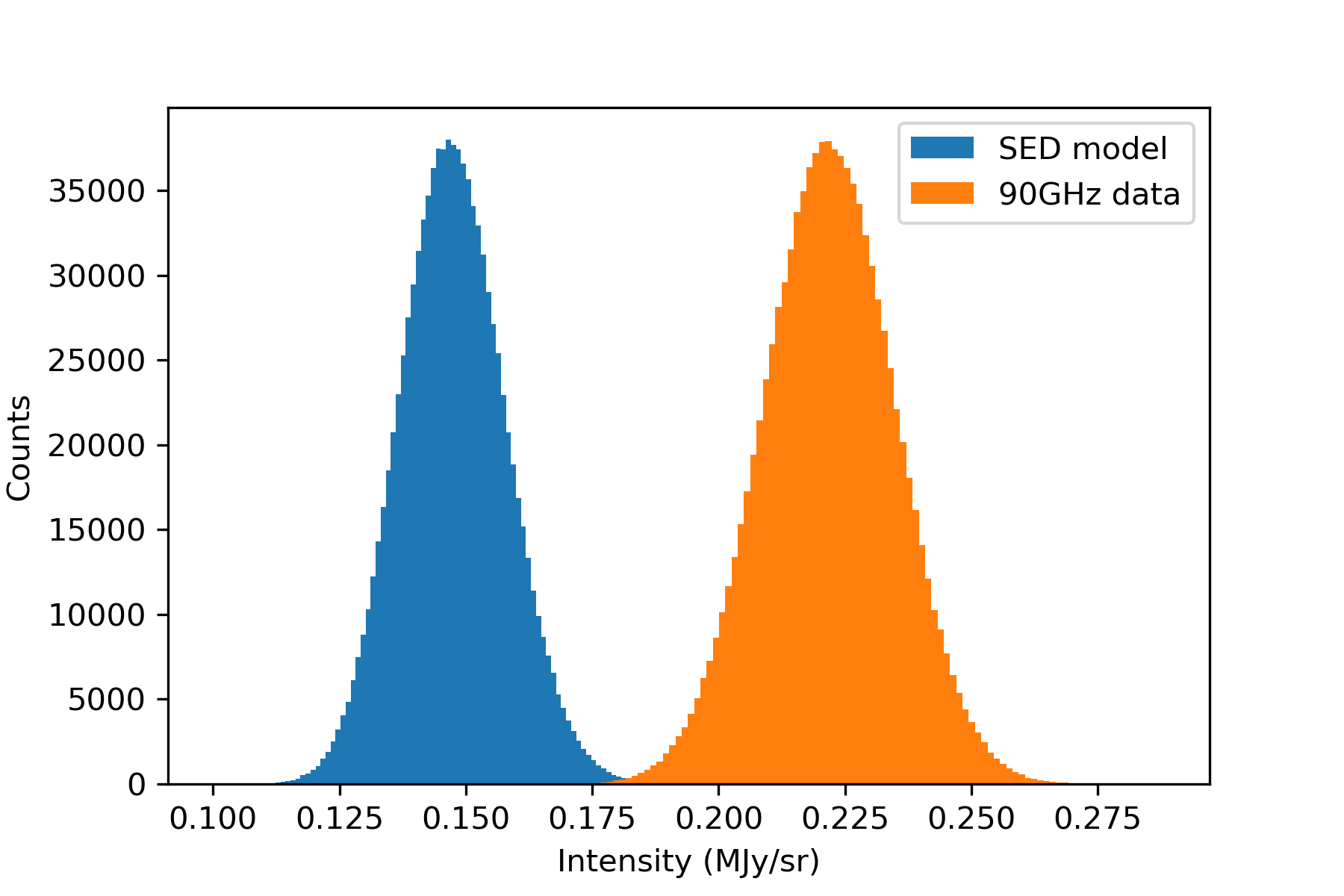}
    \caption{Example histograms from the Monte-Carlo simulations. On the left side of the graph is the range of 90~GHz fluxes predicted by the MC simulation of the \textbf{MBBno90} fit to a slice in the Serpens 1 region. The right side of the graph shows the range of values consistent with the data extracted from the 90~GHz map when map noise and calibration uncertainty are taken into consideration. As with other figures in this paper, this slice is chosen arbitrarily but is representative of the ensemble behavior.}
    \label{fig:MChistograms}
\end{figure}

\begin{figure}
    \centering
    \includegraphics[scale=0.5]{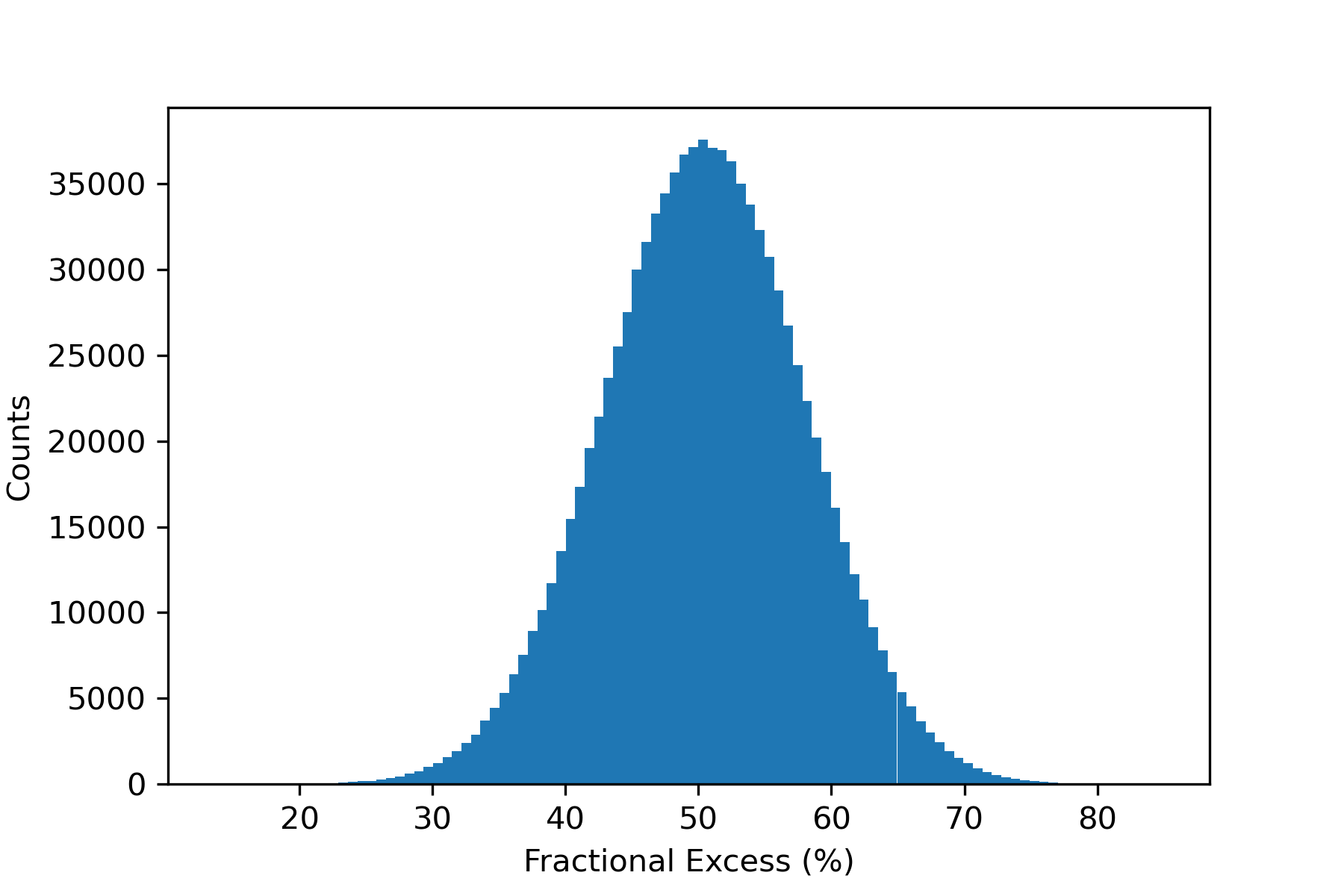}
    \caption{Distribution of fractional excess emission relative to the model as predicted by the simulations in Figure \ref{fig:MChistograms}. In this particular slice the mean excess is $\sim$50\% and is quite inconsistent with zero.}
    \label{fig:MCfracdif}
\end{figure}

\begin{figure}
    \centering
    \includegraphics[scale=0.55]{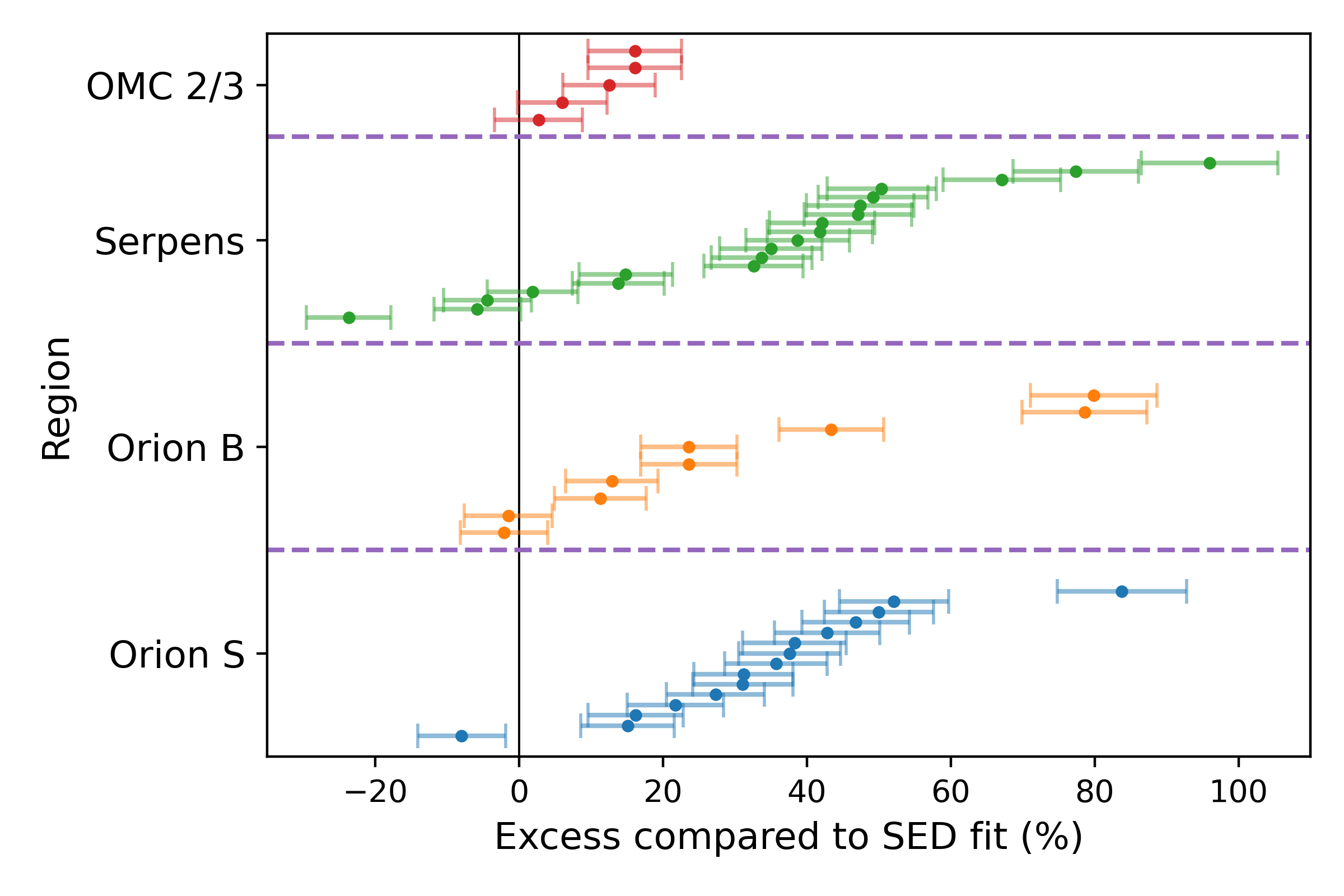}
    \caption{Visualization of the 90~GHz residuals by region with the associated 1-$\sigma$ uncertainty in the \%-excess emission relative to the model predictions. Two slices in the Orion B and one in the Orion S region are not displayed on this plot in order to preserve the readability. These slices display large excess emission values ranging from 147-307\%.}
    \label{fig:histogram_of_resid}
\end{figure}

\subsection{Overall results}

In this section, we describe the trends seen across all regions for the \textbf{MBBall} fit, \textbf{MBBno90} fit and extrapolated \textbf{MBBno90} fit. For the overall \textbf{MBBall} fit behavior we saw a median $\chi^2_n$ = 6.19, enhanced emission in 45/51 slices, and a mean elevation of the 90~GHz data of 21.9$\%$. When we restricted the SED to $\lambda \leq 2$~mm the average quality of the fit was dramatically improved, with a median $\chi^2_n$ = 1.57. The \textbf{MBBno90} fit was then extrapolated down to the 90~GHz data, which resulted in a reduced quality of the fits versus both previous analyses with a median $\chi^2_n$ = 9.96 and enhanced emission in 45/51 slices. Figure \ref{fig:meanexcesses} shows the per-slice residuals for the no90 model and, in particular, the residuals at 90 GHz are clear outliers compared to all other frequencies.

To explore this spread further and understand the significance of the result, we present the results of the Monte-Carlo noise and calibration error simulations for the 90~GHz values in particular in Figures \ref{fig:MChistograms} and \ref{fig:MCfracdif}. The two distributions from Figure \ref{fig:MChistograms} are subtracted from each other and scaled by the mean value of the model prediction to create the distribution shown in Figure \ref{fig:MCfracdif}. The mean and standard deviation of this new distribution are then extracted and used to calculate the tension with the null hypothesis of 0 excess emission at 90~GHz, which are shown as the ``Excess $\sigma$" column for each slice in Tables \ref{tab:chi-squared-serpens} and \ref{tab:chi-squared-orion}. The results sorted by region and value, along with their associated 1-$\sigma$ uncertainties are displayed in Figure \ref{fig:histogram_of_resid}. We find that, at the 4-$\sigma$ or greater significance level, nearly 60\% (30/51) of the slices are inconsistent with the modified blackbody model, which is in otherwise good agreement with the rest of our spectral bands. These probability distributions based on MC noise simulations provide clear evidence that the excesses seen in the measured data points are a real and significant effect at 90~GHz.

In addition to the above, we also examined the correlation between the $\beta$ and $T_d$ values versus the excess seen at 90~GHz. We find a correlation r of 0.133 for $\beta$ and the excess and a value of r=-0.010 for $T_d$ and the excess, both of which imply a lack of correlation between the variable. Thus, this analysis shows that there exists a significant and systematic elevation of the emission at 90~GHz on 120\arcsec scales with respect to the standard modified blackbody models that is inconsistent with both random fluctuations and calibration errors and is uncorrelated with the fit parameters.


\subsection{Two-component model}
\label{subsec:2comp}
\begin{figure}
    \centering
    \includegraphics[scale=0.5]{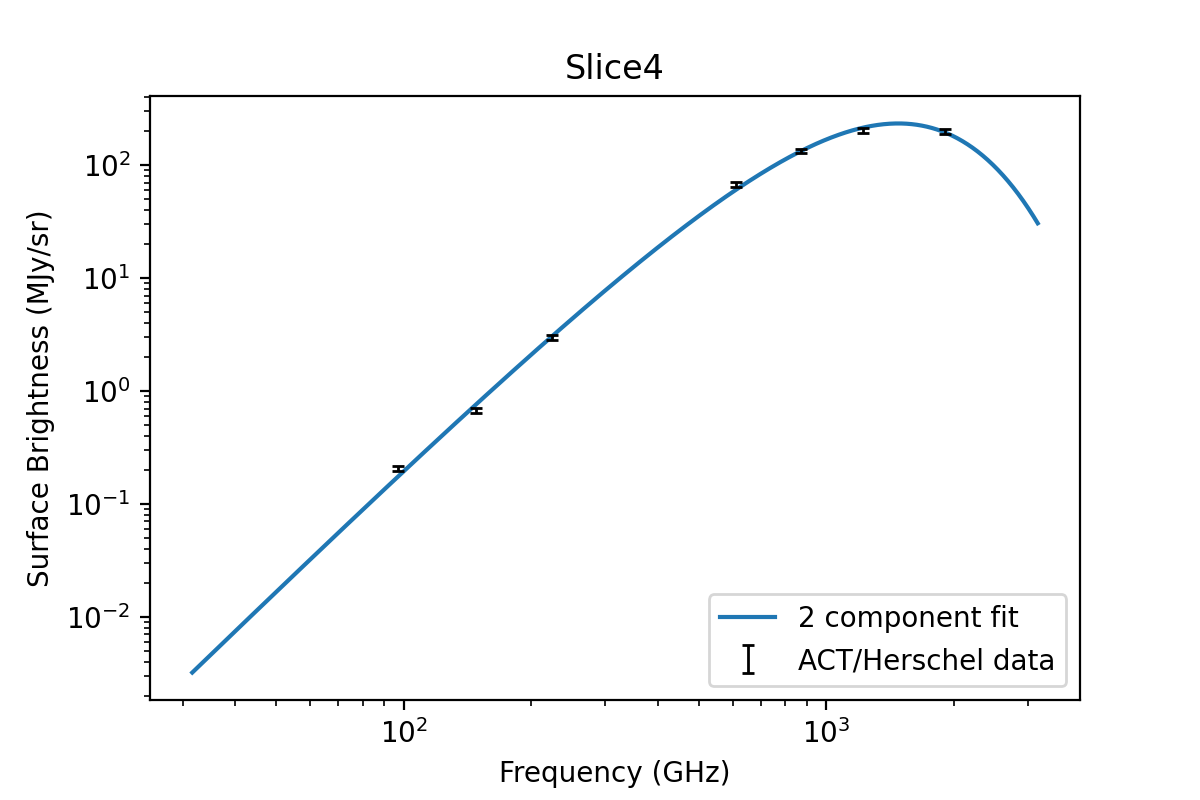}
    \caption{An example of the two-component dust model fit. Overall the fit provides a reasonable estimate of many flux density measurements, but the constraint set by the 90~GHz data causes it to over-predict the 150~GHz and still under-predict the 90~GHz values. The slice shown here is arbitrary, but represents the sample well.}
    \label{fig:2compfit}
\end{figure}

Our first additional model to consider is the two-component model (see Section \ref{sec:furthermodels}) that was described in M15. An example of this fit is shown in Figure \ref{fig:2compfit}. At 90GHz, the fits produced a lower residual than the generic modified blackbody with a mean excess of 17.0\% across all regions. The high-frequency end of the spectrum, however, was not as well fit, which led to an overall median $\chi^2_n$ = 5.51 for n=2 degrees of freedom, a minimum of $\chi^2_n$ = 0.20, and a maximum of $\chi^2_n$ = 42.54. As it stands, this model seems to provide a somewhat reasonable description of the SEDs, which is to be expected for a model with more independent parameters. Unfortunately, while it does better than the single modified blackbody at 90~GHz, it performs worse over the entire spectrum on average while still under-predicting the emission at 90~GHz. In addition, the model predicts a second component with unusually high $\beta$s (mean $\sim$2.7) and low temperatures (9-11~K) inconsistent with both the expectation of a warm second component and the NH$_3$ temperatures seen in OMC 2/3 \citep{Friesen_2017}. As it stands, this model, while reasonably motivated for the cold, high Galactic latitude cirrus, does not provide an adequate description of the emission within these warm and dense star-forming regions where less spectral coverage is available at high-resolution to balance the number of required parameters.

\begin{figure}
    \centering
    \includegraphics[scale=0.5]{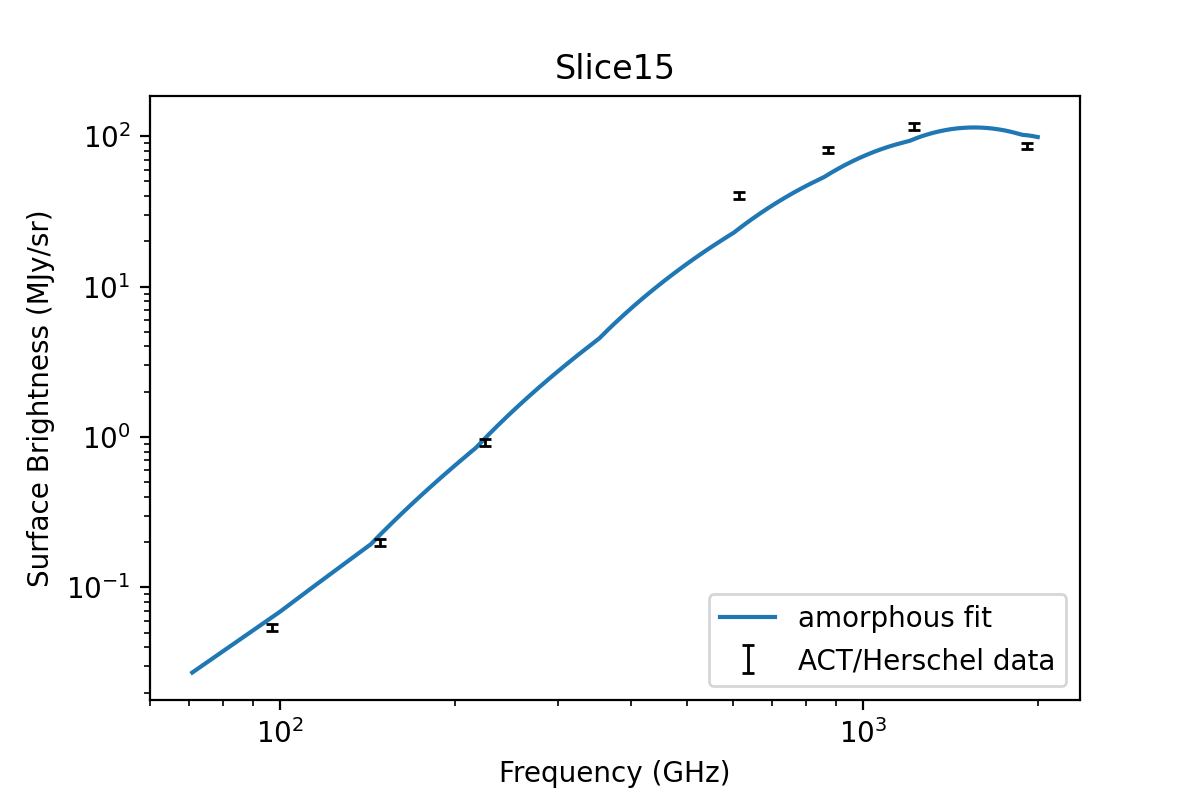}
    \caption{Example fit using the amorphous dust model on a slice SED. In this slice, the model over-predicts the 90~GHz emission while under-predicting the 250-500$\mu m$ emission, leaving the errors spread out over the SED rather than centralized at 90~GHz as with the MBB.}
    \label{fig:amorphfit}
\end{figure}


\subsection{Amorphous dust model}
\label{subsec:amorph}
We finally consider the results of fitting the amorphous grain dust model.  This model relies on a standard blackbody with an emissivity term that is a function of temperature and frequency. For our fits, we allowed the dust temperature and the normalization to be free parameters while fixing the emissivity to scale as the 17.5~K measurements from \citet{Paradis_2011}, the closest match to the mean 15.2~K temperatures we found to fit the SEDs. An example fit is shown in Figure \ref{fig:amorphfit}. This model provided a significantly worse fit to the data than the two-component model with a median $\chi^2_n$ = 8.13 for n=5 degrees of freedom, a minimum of $\chi^2_n$ = 0.37, and a maximum of $\chi^2_n$ = 115.1. Interestingly, this is the sole model which displays large errors in predictions across the spectrum, rather than failing at any one specific frequency repeatedly. This model fares better than any other model at predicting the surface brightness with an average excess of -1.04\%, meaning it slightly over predicted the 90~GHz flux on average. The overall quality of this fit, however, is bad. We find that is it comparable to that of the 90~GHz constrained \textbf{MBBall} fit and is significantly worse than that of the short wavelength \textbf{MBBno90} fit. There exists hope for this model, however, as more detailed $\epsilon(T,\nu)$ data would allow for us to include the $\epsilon$-temperature in the model in the future.
\section{discussion and Conclusions}
\label{sec:concs}

Here we have presented a new study with significant evidence for enhanced emission, relative to a modified blackbody model, at 90~GHz in the nearby molecular cloud regions of Orion A, Orion B, and Serpens at 120\arcsec resolution. Considering first the analysis of only the OMC 2/3 region at 120\arcsec resolution and the search for evidence of enhanced emission using the ACT and Herschel data we found evidence for enhanced emission at 90~GHz, with all slices showing enhanced emission above the MBB. The data as compared to the \textbf{MBBall} fit model has a modest 4.9\% average elevation and when compared to the extrapolated model it rose to 10.9\%, though no slices within this region were inconsistent with 0 excess to our 4-$\sigma$ cutoff. In this region we also saw that \textit{no90 fit} produced the best average agreement with the model and that fit, when extrapolated to 90~GHz, significantly degraded the quality of the fit relative to both other methods.

With the prototype analysis of the OMC 2/3 region completed and showing tentative evidence for emission above an MBB spectrum, we extended our study to other molecular clouds within the HGBS. We chose to focus on three targets at similar distances ($\sim$400pc) to  the OMC 2/3 filament, the Orion A-S, Orion B, and Serpens regions. Within these regions we found an even higher fraction (41 of 46) of the slices showing elevated emission and a striking departure in the behavior of the 90~GHz emission, with 30/46 slices displaying a 4-$\sigma$ or greater inconsistency with the modified blackbody at 90G~Hz. These regions have provided much stronger evidence for enhanced emission than even the OMC 2/3 region which prompted this study and was the backbone of the results in M20.

In addition to the results of the modified blackbody SED models, we have considered the possible effects of free-free and line emission contamination as well as two-component and amorphous dust models to explain the 90~GHz discrepancies. In the case of  free-free contamination, we found only two slices (one of which was selected due to a low-frequency source for code testing) out of the 53 total which were identified as co-located with an HII region from the NVSS+WISE catalogs. We elected to remove these slices from consideration due to the likely presence of contamination. For the case of contamination by molecular line emission, we found that the Planck data were indeed contaminated by the CO line.  The Planck beam was large compared to our slices, however,  so the CO line emission appeared as an approximately constant background term which was removed via the deramping. The results were in line with the ACT only maps. The other potential cause of contamination is the HCN line, which we can ignore given how weak the emission was.  Turning now to the additional models, we found that, while they resulted in similar reduced $\chi_n^2$ values as the \textbf{MBBall} fit model, they each produced results which were concerning, either unreasonable parameters, poor fits, or both. In the case of the two-component model, we restricted ourselves to a fixed cold component model from M15 in combination with a free, warm component. This resulted in overall slightly improved (but still poor) fits from the \textbf{MBBall} fit ($\chi_n^2$ = 5.51 vs $\chi_n^2$ = 6.19), a systematic under-prediction of the 90~GHz flux (17.0\% low), and temperatures inconsistent with the NH$_3$ in OMC 2/3 and $\beta$s averaging over 2.6, which are unusually high. On the other hand, the amorphous dust grain model provided a drastically worse fit to the curves ($\chi_n^2$ = 8.81 vs $\chi_n^2$ = 6.19) than the \textbf{MBBall} fit and good agreement with the model at 90~GHz (1.04\% over prediction). However, this model fails to describe the high-frequency end of the spectrum well, as is particularly evident in Figure \ref{fig:amorphfit}. Both of these models have the potential to provide a better description of the SED than a generic modified blackbody; however, in the case of the amorphous dust, more measurements of the behavior at different temperatures would be required, and for the two-component model, we would need more spectral coverage to comfortably use a six-parameter model to fit the data.

When compared to the results of M20, we found that our data reduction pipeline produced results in excellent agreement with M20, having a median difference and standard deviation of $\Delta T_d/\sigma_{T_d}$ = -0.190$\pm 0.095$ and $\Delta \beta/\sigma_{\beta}$ = 0.13$\pm 0.11$. The method that we used in this paper has the benefits of being computationally lightweight and having the ability to extract information on these clouds solely through existing survey and archival data. This study has shown that our pipeline is a viable method through which we can combine the data at 120\arcsec resolution and measure the SEDs across a variety of sources of varying brightness. By extracting information about the spectral energy distribution in these regions, and especially about the prevalence of enhancement at 90~GHz, we have paved the way for complementary observations and analyses to examine the emission on smaller angular scales and lower frequencies. 

With our concerns about contamination assuaged and additional models fit and ruled out, we consider possible reasons for this deviation from the modified blackbody. While we are not sure what the source of this emission is, whether it is dust-based or from some other source, analyses have so far ruled out contributions from spectral line emission of CO and HCN, AME, and spinning dust variants. This leaves a gap in our understanding of the dust,  though we should not be surprised that such a simple MBB model fails to encapsulate the complex behavior of the emission from these clouds. It is known already that low-frequency emission does not extrapolate well to higher frequencies \citep{planck2015_X}, and perhaps the spectral range we consider is simply too wide for a single temperature or fixed spectral index modified blackbody to accurately describe. This seemingly common elevation of emission at 90~GHz is troubling, as it means that using standard dust models for core envelopes (e.g., \citet{Ossenkopf}) to convert thermal dust emission at 3mm to dust masses may overestimate the measured masses significantly. Whatever the source of this emission may be, it represents a significant deviation from the generic model and therefore warrants further investigation.

In order to better understand the properties of this enhanced emission, there are three obvious paths for future investigation. The first path is to consider these regions in polarization, which would trace the dust emission in particular and could not be contaminated by free-free or other unpolarized emission. This would give us an understanding of the association with dust or a alternative contributing source. This would require sensitive surveys of the molecular clouds in polarization, on well-matched angular scales to the currently existing ACT polarization data. The second is to further extend this study to additional clouds and regions. We have so far concerned ourselves with only a small fraction of the available HGBS and ACT data. There exist data on tens of clouds available to study with a wealth of information about this behavior to extract. Our analysis has shown a systematic increase in the surface brightness at 90~GHz, with more than 90$\%$ of the $>50$ slices showing an increase even when the model is constrained by that data. We have studied only 3 relatively small regions within these cloud, however, leaving potentially hundreds of slices for future analyses. These future works would provide significant spatial coverage to test the ubiquity of the 90~GHz excess, and a much larger sample size. The third path is to delve further into the behavior of the clouds studied in this paper with the methods of M20 and data available at higher resolution. We have provided evidence that there exists enhanced emission in these regions for 120\arcsec resolution at 90~GHz, which makes them prime targets for followup observations with high resolution on instruments such as MUSTANG2 and the GBT Ka-band receiver, the latter of which would provide the very low frequency coverage that showed the largest enhancements in M20. Additionally, where legacy spectral coverage of these clouds lapses, there exist instruments such as NIKA2 \citep{Calvo_2016}, at 1.25 and 2~mm, on the IRAM 30m telescope which provide high angular resolution coverage in the portion spectrum between the GBT (3~mm) instruments and the Herschel data (160-500$\mu$m), crucial for bridging the gap in the tail of the SED between 600~GHz and 90~GHz and determining the presence of enhanced emission. With this suite of available instruments, there exists significant opportunity for followup observation and analysis of these regions to determine the nature of dust emission in star-forming molecular clouds.

\textbf{Acknowledgements}: IL would like to acknowledge the assistance of Marius Lungu from the ACT collaboration in finding and sharing the ACT bandpass data used within this paper. This project has received funding from the European Research Council (ERC) under the European Union’s Horizon 2020 research and innovation programme (Grant agreement No. 851435). \textbf{The National Radio Astronomy Observatory is a facility of the National Science Foundation operated under cooperative agreement by Associated Universities, Inc. the authors would also like to acknowledge the thoughtful and constructive suggestions of the referee which helped to improve the manuscript.}

\textit{Software}: Astropy \citep{astropy1,astropy2}, SciPy \citep{2020SciPy}, Matplotlib \citep{Hunter:2007}, DS9 \citep{ds9}

\bibliographystyle{aasjournal}
\bibliography{References}

\begin{thebibliography}{}
\expandafter\ifx\csname natexlab\endcsname\relax\def\natexlab#1{#1}\fi
\providecommand{\url}[1]{\href{#1}{#1}}
\providecommand{\dodoi}[1]{doi:~\href{http://doi.org/#1}{\nolinkurl{#1}}}
\providecommand{\doeprint}[1]{\href{http://ascl.net/#1}{\nolinkurl{http://ascl.net/#1}}}
\providecommand{\doarXiv}[1]{\href{https://arxiv.org/abs/#1}{\nolinkurl{https://arxiv.org/abs/#1}}}

\bibitem[{Adam {et~al.}(2016)Adam, Ade, Aghanim, Alves, Arnaud, Ashdown,
  Aumont, Baccigalupi, Banday, \& et~al.}]{planck2015_X}
Adam, R., Ade, P. A.~R., Aghanim, N., {et~al.} 2016, Astronomy \& Astrophysics,
  594, A10, \dodoi{10.1051/0004-6361/201525967}

\bibitem[{Ade {et~al.}(2014{\natexlab{a}})Ade, Aghanim, Alves, Armitage-Caplan,
  Arnaud, Ashdown, Atrio-Barandela, Aumont, Aussel, \& et~al.}]{Planck2013_1}
Ade, P. A.~R., Aghanim, N., Alves, M. I.~R., {et~al.} 2014{\natexlab{a}},
  Astronomy \& Astrophysics, 571, A1, \dodoi{10.1051/0004-6361/201321529}

\bibitem[{Ade {et~al.}(2014{\natexlab{b}})Ade, Aghanim, Armitage-Caplan,
  Arnaud, Ashdown, Atrio-Barandela, Aumont, Baccigalupi, Banday, \&
  et~al.}]{planck2014_7}
Ade, P. A.~R., Aghanim, N., Armitage-Caplan, C., {et~al.} 2014{\natexlab{b}},
  Astronomy \& Astrophysics, 571, A7, \dodoi{10.1051/0004-6361/201321535}

\bibitem[{Ade {et~al.}(2014{\natexlab{c}})Ade, Aghanim, Alves, Armitage-Caplan,
  Arnaud, Ashdown, Atrio-Barandela, Aumont, Baccigalupi, \&
  et~al.}]{planck2014_13}
Ade, P. A.~R., Aghanim, N., Alves, M. I.~R., {et~al.} 2014{\natexlab{c}},
  Astronomy \& Astrophysics, 571, A13, \dodoi{10.1051/0004-6361/201321553}

\bibitem[{Aghanim {et~al.}(2016)Aghanim, Ashdown, Aumont, Baccigalupi,
  Ballardini, Banday, Barreiro, Bartolo, Basak, \& et~al.}]{planck_int_xlviii}
Aghanim, N., Ashdown, M., Aumont, J., {et~al.} 2016, Astronomy \& Astrophysics,
  596, A109, \dodoi{10.1051/0004-6361/201629022}

\bibitem[{Anderson {et~al.}(2014)Anderson, Bania, Balser, Cunningham, Wenger,
  Johnstone, \& Armentrout}]{Anderson_2014}
Anderson, L.~D., Bania, T.~M., Balser, D.~S., {et~al.} 2014, The Astrophysical
  Journal Supplement Series, 212, 1, \dodoi{10.1088/0067-0049/212/1/1}

\bibitem[{André {et~al.}(2014)André, Di~Francesco, Ward-Thompson, Inutsuka,
  Pudritz, \& Pineda}]{Andre_2014}
André, P., Di~Francesco, J., Ward-Thompson, D., {et~al.} 2014, Protostars and
  Planets VI, \dodoi{10.2458/azu_uapress_9780816531240-ch002}

\bibitem[{Arzoumanian {et~al.}(2019)Arzoumanian, André, Könyves, Palmeirim,
  Roy, Schneider, Benedettini, Didelon, Di~Francesco, Kirk, \&
  et~al.}]{Arzoumanian_2019}
Arzoumanian, D., André, P., Könyves, V., {et~al.} 2019, Astronomy \&
  Astrophysics, 621, A42, \dodoi{10.1051/0004-6361/201832725}

\bibitem[{Bendo {et~al.}(2013)Bendo, Griffin, Bock, Conversi, Dowell, Lim, Lu,
  North, Papageorgiou, Pearson, \& et~al.}]{Bendo_2013}
Bendo, G.~J., Griffin, M.~J., Bock, J.~J., {et~al.} 2013, Monthly Notices of
  the Royal Astronomical Society, 433, 3062–3078,
  \dodoi{10.1093/mnras/stt948}

\bibitem[{Bertoldi {et~al.}(2003)Bertoldi, Carilli, Cox, Fan, Strauss, Beelen,
  Omont, \& Zylka}]{Bertoldi_2003}
Bertoldi, F., Carilli, C.~L., Cox, P., {et~al.} 2003, Astronomy \&
  Astrophysics, 406, L55–L58, \dodoi{10.1051/0004-6361:20030710}

\bibitem[{Calvo {et~al.}(2016)Calvo, Benoît, Catalano, Goupy, Monfardini,
  Ponthieu, Barria, Bres, Grollier, Garde, \& et~al.}]{Calvo_2016}
Calvo, M., Benoît, A., Catalano, A., {et~al.} 2016, Journal of Low Temperature
  Physics, 184, 816–823, \dodoi{10.1007/s10909-016-1582-0}

\bibitem[{Chen {et~al.}(2019)Chen, King, Li, Fissel, \& Mazzei}]{Chen_2019}
Chen, C.-Y., King, P.~K., Li, Z.-Y., Fissel, L.~M., \& Mazzei, R.~R. 2019,
  Monthly Notices of the Royal Astronomical Society, 485, 3499–3513,
  \dodoi{10.1093/mnras/stz618}

\bibitem[{Condon {et~al.}(1998)Condon, Cotton, Greisen, Yin, Perley, Taylor, \&
  Broderick}]{Condon_1998}
Condon, J.~J., Cotton, W.~D., Greisen, E.~W., {et~al.} 1998, The Astronomical
  Journal, 115, 1693, \dodoi{10.1086/300337}

\bibitem[{Coupeaud {et~al.}(2011)Coupeaud, Demyk, Meny, Nayral, Delpech,
  Leroux, Depecker, Creff, Brubach, \& Roy}]{Coupeaud_2011}
Coupeaud, A., Demyk, K., Meny, C., {et~al.} 2011, Astronomy \& Astrophysics,
  535, A124, \dodoi{10.1051/0004-6361/201116945}

\bibitem[{Crutcher(2012)}]{crutcher_magnetic}
Crutcher, R.~M. 2012, Annual Review of Astronomy and Astrophysics, 50, 29,
  \dodoi{10.1146/annurev-astro-081811-125514}

\bibitem[{{Dame} {et~al.}(2001){Dame}, {Hartmann}, \&
  {Thaddeus}}]{dame_co_orion}
{Dame}, T.~M., {Hartmann}, D., \& {Thaddeus}, P. 2001, \apj, 547, 792,
  \dodoi{10.1086/318388}

\bibitem[{{Dicker} {et~al.}(2014){Dicker}, {Ade}, {Aguirre}, {Brevik}, {Cho},
  {Datta}, {Devlin}, {Dober}, {Egan}, {Ford}, {Ford}, {Hilton}, {Irwin},
  {Mason}, {Marganian}, {Mello}, {McMahon}, {Mroczkowski}, {Rosenman},
  {Tucker}, {Vale}, {White}, {Whitehead}, \& {Young}}]{dicker_2014}
{Dicker}, S.~R., {Ade}, P.~A.~R., {Aguirre}, J., {et~al.} 2014, Journal of Low
  Temperature Physics, 176, 808, \dodoi{10.1007/s10909-013-1070-8}

\bibitem[{Draine \& Hensley(2013)}]{2013_draine+hensley}
Draine, B.~T., \& Hensley, B. 2013, The Astrophysical Journal, 765, 159,
  \dodoi{10.1088/0004-637x/765/2/159}

\bibitem[{Draine \& Lazarian(1999)}]{Draine_1999}
Draine, B.~T., \& Lazarian, A. 1999, The Astrophysical Journal, 512, 740–754,
  \dodoi{10.1086/306809}

\bibitem[{Eales {et~al.}(2012)Eales, Smith, Auld, Baes, Bendo, Bianchi,
  Boselli, Ciesla, Clements, Cooray, Cortese, Davies, De~Looze, Galametz, Gear,
  Gentile, Gomez, Fritz, Hughes, \& Wilson}]{eales_2012}
Eales, S., Smith, M., Auld, R., {et~al.} 2012, The Astrophysical Journal, 761,
  168, \dodoi{10.1088/0004-637X/761/2/168}

\bibitem[{Fiorellino {et~al.}(2020)Fiorellino, Elia, André, Men’shchikov,
  Pezzuto, Schisano, Könyves, Arzoumanian, Benedettini, Ward-Thompson, \&
  et~al.}]{Fiorellino_2020}
Fiorellino, E., Elia, D., André, P., {et~al.} 2020, Monthly Notices of the
  Royal Astronomical Society, 500, 4257–4276, \dodoi{10.1093/mnras/staa3420}

\bibitem[{Fissel {et~al.}(2019)Fissel, Ade, Angilè, Ashton, Benton, Chen,
  Cunningham, Devlin, Dober, Friesen, \& et~al.}]{Fissel_2019}
Fissel, L.~M., Ade, P. A.~R., Angilè, F.~E., {et~al.} 2019, The Astrophysical
  Journal, 878, 110, \dodoi{10.3847/1538-4357/ab1eb0}

\bibitem[{Friesen {et~al.}(2017)Friesen, Pineda, Rosolowsky, Alves,
  Chacón-Tanarro, Chen, Chen, Di~Francesco, Keown, Kirk, \&
  et~al.}]{Friesen_2017}
Friesen, R.~K., Pineda, J.~E., Rosolowsky, E., {et~al.} 2017, The Astrophysical
  Journal, 843, 63, \dodoi{10.3847/1538-4357/aa6d58}

\bibitem[{Groves {et~al.}(2015)Groves, Schinnerer, Leroy, Galametz, Walter,
  Bolatto, Hunt, Dale, Calzetti, Croxall, \& et~al.}]{Groves_2015}
Groves, B.~A., Schinnerer, E., Leroy, A., {et~al.} 2015, The Astrophysical
  Journal, 799, 96, \dodoi{10.1088/0004-637x/799/1/96}

\bibitem[{Heitsch {et~al.}(2001)Heitsch, Zweibel, Mac~Low, Li, \&
  Norman}]{Heitsch_2001}
Heitsch, F., Zweibel, E.~G., Mac~Low, M., Li, P., \& Norman, M.~L. 2001, The
  Astrophysical Journal, 561, 800–814, \dodoi{10.1086/323489}

\bibitem[{Henderson {et~al.}(2016)Henderson, Allison, Austermann, Baildon,
  Battaglia, Beall, Becker, De~Bernardis, Bond, Calabrese, \&
  et~al.}]{Henderson_2016}
Henderson, S.~W., Allison, R., Austermann, J., {et~al.} 2016, Journal of Low
  Temperature Physics, 184, 772–779, \dodoi{10.1007/s10909-016-1575-z}

\bibitem[{Hennebelle \& Falgarone(2012)}]{Hennebelle_2012}
Hennebelle, P., \& Falgarone, E. 2012, The Astronomy and Astrophysics Review,
  20, \dodoi{10.1007/s00159-012-0055-y}

\bibitem[{Hensley \& Bull(2018)}]{Hensley_2018}
Hensley, B.~S., \& Bull, P. 2018, The Astrophysical Journal, 853, 127,
  \dodoi{10.3847/1538-4357/aaa489}

\bibitem[{Holland {et~al.}(2013)Holland, Bintley, Chapin, Chrysostomou, Davis,
  Dempsey, Duncan, Fich, Friberg, Halpern, \& et~al.}]{Holland_2013}
Holland, W.~S., Bintley, D., Chapin, E.~L., {et~al.} 2013, Monthly Notices of
  the Royal Astronomical Society, 430, 2513–2533,
  \dodoi{10.1093/mnras/sts612}

\bibitem[{Hunter(2007)}]{Hunter:2007}
Hunter, J.~D. 2007, Computing in Science \& Engineering, 9, 90,
  \dodoi{10.1109/MCSE.2007.55}

\bibitem[{{Joye} \& {Mandel}(2003)}]{ds9}
{Joye}, W.~A., \& {Mandel}, E. 2003, in Astronomical Society of the Pacific
  Conference Series, Vol. 295, Astronomical Data Analysis Software and Systems
  XII, ed. H.~E. {Payne}, R.~I. {Jedrzejewski}, \& R.~N. {Hook}, 489

\bibitem[{Könyves {et~al.}(2020)Könyves, André, Arzoumanian, Schneider,
  Men’shchikov, Bontemps, Ladjelate, Didelon, Pezzuto, Benedettini, \&
  et~al.}]{Konyves_2020}
Könyves, V., André, P., Arzoumanian, D., {et~al.} 2020, Astronomy \&
  Astrophysics, 635, A34, \dodoi{10.1051/0004-6361/201834753}

\bibitem[{Mason {et~al.}(2020)Mason, Dicker, Sadavoy, Stanchfield, Mroczkowski,
  Romero, Friesen, Sarazin, Sievers, Stanke, \& et~al.}]{Mason_2020}
Mason, B., Dicker, S., Sadavoy, S., {et~al.} 2020, The Astrophysical Journal,
  893, 13, \dodoi{10.3847/1538-4357/ab734a}

\bibitem[{{Meisner} \& {Finkbeiner}(2015)}]{Meisner2015}
{Meisner}, A.~M., \& {Finkbeiner}, D.~P. 2015, \apj, 798, 88,
  \dodoi{10.1088/0004-637X/798/2/88}

\bibitem[{Meny {et~al.}(2007)Meny, Gromov, Boudet, Bernard, Paradis, \&
  Nayral}]{Meny_2007}
Meny, C., Gromov, V., Boudet, N., {et~al.} 2007, Astronomy \& Astrophysics,
  468, 171–188, \dodoi{10.1051/0004-6361:20065771}

\bibitem[{Naess {et~al.}(2020)Naess, Aiola, Austermann, Battaglia, Beall,
  Becker, Bond, Calabrese, Choi, Cothard, \& et~al.}]{Naess_2020}
Naess, S., Aiola, S., Austermann, J.~E., {et~al.} 2020, Journal of Cosmology
  and Astroparticle Physics, 2020, 046–046,
  \dodoi{10.1088/1475-7516/2020/12/046}

\bibitem[{{Nashimoto} {et~al.}(2020){Nashimoto}, {Hattori}, {Poidevin}, \&
  {G{\'e}nova-Santos}}]{nashimoto2020}
{Nashimoto}, M., {Hattori}, M., {Poidevin}, F., \& {G{\'e}nova-Santos}, R.
  2020, \apjl, 900, L40, \dodoi{10.3847/2041-8213/abb29d}

\bibitem[{{Ohashi} {et~al.}(2014){Ohashi}, {Tatematsu}, {Choi}, {Kang},
  {Umemoto}, {Lee}, {Hirota}, {Yamamoto}, \& {Mizuno}}]{OrionA-HCN}
{Ohashi}, S., {Tatematsu}, K., {Choi}, M., {et~al.} 2014, \pasj, 66, 119,
  \dodoi{10.1093/pasj/psu116}

\bibitem[{{Ossenkopf} \& {Henning}(1994)}]{Ossenkopf}
{Ossenkopf}, V., \& {Henning}, T. 1994, \aap, 291, 943

\bibitem[{Paradis {et~al.}(2011)Paradis, Bernard, Mény, \&
  Gromov}]{Paradis_2011}
Paradis, D., Bernard, J.-P., Mény, C., \& Gromov, V. 2011, Astronomy \&
  Astrophysics, 534, A118, \dodoi{10.1051/0004-6361/201116862}

\bibitem[{Paradis {et~al.}(2019)Paradis, Mény, Juvela, Noriega-Crespo, \&
  Ristorcelli}]{Paradis_2019}
Paradis, D., Mény, C., Juvela, M., Noriega-Crespo, A., \& Ristorcelli, I.
  2019, Astronomy \& Astrophysics, 627, A15,
  \dodoi{10.1051/0004-6361/201935158}

\bibitem[{Peretto {et~al.}(2013)Peretto, Fuller, Duarte-Cabral, Avison,
  Hennebelle, Pineda, André, Bontemps, Motte, Schneider, \&
  et~al.}]{Peretto_2013}
Peretto, N., Fuller, G.~A., Duarte-Cabral, A., {et~al.} 2013, Astronomy \&
  Astrophysics, 555, A112, \dodoi{10.1051/0004-6361/201321318}

\bibitem[{Pilbratt {et~al.}(2010)Pilbratt, Riedinger, Passvogel, Crone, Doyle,
  Gageur, Heras, Jewell, Metcalfe, Ott, \& et~al.}]{Pilbratt_2010}
Pilbratt, G.~L., Riedinger, J.~R., Passvogel, T., {et~al.} 2010, Astronomy and
  Astrophysics, 518, L1, \dodoi{10.1051/0004-6361/201014759}

\bibitem[{Polychroni {et~al.}(2013)Polychroni, Schisano, Elia, Roy, Molinari,
  Martin, André, Turrini, Rygl, Di~Francesco, \& et~al.}]{Polychroni_2013}
Polychroni, D., Schisano, E., Elia, D., {et~al.} 2013, The Astrophysical
  Journal, 777, L33, \dodoi{10.1088/2041-8205/777/2/l33}

\bibitem[{Price-Whelan {et~al.}(2018)Price-Whelan, Sipőcz, Günther, Lim,
  Crawford, Conseil, Shupe, Craig, Dencheva, \& et~al.}]{astropy2}
Price-Whelan, A.~M., Sipőcz, B.~M., Günther, H.~M., {et~al.} 2018, The
  Astronomical Journal, 156, 123, \dodoi{10.3847/1538-3881/aabc4f}

\bibitem[{Robitaille {et~al.}(2013)Robitaille, Tollerud, Greenfield,
  Droettboom, Bray, Aldcroft, Davis, Ginsburg, Price-Whelan, \&
  et~al.}]{astropy1}
Robitaille, T.~P., Tollerud, E.~J., Greenfield, P., {et~al.} 2013, Astronomy \&
  Astrophysics, 558, A33, \dodoi{10.1051/0004-6361/201322068}

\bibitem[{{Rydbeck} {et~al.}(1981){Rydbeck}, {Hjalmarson}, {Rydbeck}, {Ellder},
  {Olofsson}, \& {Sume}}]{1981_HCN}
{Rydbeck}, O.~E.~H., {Hjalmarson}, A., {Rydbeck}, G., {et~al.} 1981, \apjl,
  243, L41, \dodoi{10.1086/183439}

\bibitem[{Sadavoy {et~al.}(2016)Sadavoy, Stutz, Schnee, Mason, Di~Francesco, \&
  Friesen}]{Sadavoy_2016}
Sadavoy, S.~I., Stutz, A.~M., Schnee, S., {et~al.} 2016, Astronomy \&
  Astrophysics, 588, A30, \dodoi{10.1051/0004-6361/201527364}

\bibitem[{Sadavoy {et~al.}(2013)Sadavoy, Di~Francesco, Johnstone, Currie,
  Drabek, Hatchell, Nutter, André, Arzoumanian, Benedettini, \&
  et~al.}]{Sadavoy_2013}
Sadavoy, S.~I., Di~Francesco, J., Johnstone, D., {et~al.} 2013, The
  Astrophysical Journal, 767, 126, \dodoi{10.1088/0004-637x/767/2/126}

\bibitem[{Schnee {et~al.}(2014)Schnee, Mason, Di~Francesco, Friesen, Li,
  Sadavoy, \& Stanke}]{Schnee_2014}
Schnee, S., Mason, B., Di~Francesco, J., {et~al.} 2014, Monthly Notices of the
  Royal Astronomical Society, 444, 2303–2312, \dodoi{10.1093/mnras/stu1596}

\bibitem[{Schnee {et~al.}(2009)Schnee, Enoch, Noriega-Crespo, Sayers, Terebey,
  Caselli, Foster, Goodman, Kauffmann, Padgett, \& et~al.}]{Schnee_2009}
Schnee, S., Enoch, M., Noriega-Crespo, A., {et~al.} 2009, The Astrophysical
  Journal, 708, 127–136, \dodoi{10.1088/0004-637x/708/1/127}

\bibitem[{Shetty {et~al.}(2009)Shetty, Kauffmann, Schnee, \&
  Goodman}]{Shetty_2009}
Shetty, R., Kauffmann, J., Schnee, S., \& Goodman, A.~A. 2009, The
  Astrophysical Journal, 696, 676–680, \dodoi{10.1088/0004-637x/696/1/676}

\bibitem[{Staguhn {et~al.}(2006)Staguhn, Benford, Allen, Moseley, Sharp, Ames,
  Brunswig, Chuss, Dwek, Maher, Marx, Miller, Navarro, \&
  Wollack}]{staguhn_2006}
Staguhn, J.~G., Benford, D.~J., Allen, C.~A., {et~al.} 2006, in Millimeter and
  Submillimeter Detectors and Instrumentation for Astronomy III, ed.
  J.~Zmuidzinas, W.~S. Holland, S.~Withington, \& W.~D. Duncan, Vol. 6275,
  International Society for Optics and Photonics (SPIE), 428 -- 436,
  \dodoi{10.1117/12.671970}

\bibitem[{{Stanchfield} {et~al.}(2016){Stanchfield}, {Ade}, {Aguirre},
  {Brevik}, {Cho}, {Datta}, {Devlin}, {Dicker}, {Dober}, {Egan}, {Ford},
  {Hilton}, {Hubmayr}, {Irwin}, {Marganian}, {Mason}, {Mates}, {McMahon},
  {Mello}, {Mroczkowski}, {Romero}, {Tucker}, {Vale}, {White}, {Whitehead}, \&
  {Young}}]{stanchfield_2016}
{Stanchfield}, S.~M., {Ade}, P.~A.~R., {Aguirre}, J., {et~al.} 2016, Journal of
  Low Temperature Physics, 184, 460, \dodoi{10.1007/s10909-016-1570-4}

\bibitem[{Stutz \& Gould(2016)}]{Stutz_2016}
Stutz, A.~M., \& Gould, A. 2016, Astronomy \& Astrophysics, 590, A2,
  \dodoi{10.1051/0004-6361/201527979}

\bibitem[{Stutz \& Kainulainen(2015)}]{Stutz_2015}
Stutz, A.~M., \& Kainulainen, J. 2015, Astronomy \& Astrophysics, 577, L6,
  \dodoi{10.1051/0004-6361/201526243}

\bibitem[{Stutz {et~al.}(2013)Stutz, Tobin, Stanke, Megeath, Fischer,
  Robitaille, Henning, Ali, di~Francesco, Furlan, \& et~al.}]{Stutz_2013}
Stutz, A.~M., Tobin, J.~J., Stanke, T., {et~al.} 2013, The Astrophysical
  Journal, 767, 36, \dodoi{10.1088/0004-637x/767/1/36}

\bibitem[{Vansyngel {et~al.}(2018)Vansyngel, Boulanger, Ghosh, Wandelt, Aumont,
  Bracco, Levrier, Martin, \& Montier}]{Vansyngel_2018}
Vansyngel, F., Boulanger, F., Ghosh, T., {et~al.} 2018, Astronomy \&
  Astrophysics, 618, C4, \dodoi{10.1051/0004-6361/201629992e}

\bibitem[{Virtanen {et~al.}(2020)Virtanen, Gommers, Oliphant, Haberland, Reddy,
  Cournapeau, Burovski, Peterson, {Weckesser}, {Bright}, {van der Walt},
  {Brett}, {Wilson}, {Jarrod Millman}, {Mayorov}, {Nelson}, {Jones}, {Kern},
  {Larson}, {Carey}, {Polat}, {Feng}, {Moore}, {Vand erPlas}, {Laxalde},
  {Perktold}, {Cimrman}, {Henriksen}, {Quintero}, {Harris}, {Archibald},
  {Ribeiro}, {Pedregosa}, {van Mulbregt}, \& {Contributors}}]{2020SciPy}
Virtanen, P., Gommers, R., Oliphant, T.~E., {et~al.} 2020, Nature Methods

\end{thebibliography}

\end{document}